astro-ph/9407003 1 JUL 94

# The Ensemble Variability Properties of Faint QSOs

Dario Trevese[1]
Istituto Astronomico, Universita degli Studi "La Sapienza,"
via G.M. Lancisi 29, 00161, Rome, Italy
Electronic mail: trevese@astrom.astro.it

Richard G. Kron[1]
Fermi National Accelerator Laboratory,
MS 127, Box 500, Batavia, Illinois 60510
Electronic mail: rich@oddjob.uchicago.edu

Steven R. Majewski[1,2]
The Observatories of the Carnegie Institution of Washington,
813 Santa Barbara St., Pasadena, California 91101
Electronic mail: srm@ociw.edu

Matthew A. Bershady[1,2]
University of California Observatories / Lick Observatory,
University of California, Santa Cruz, CA 95064
Electronic mail: mab@lick.ucsc.edu

David C. Koo[1]
University of California Observatories / Lick Observatory,
Board of Studies in Astronomy and Astrophysics,
University of California, Santa Cruz, CA 95064
Electronic mail: koo@lick.ucsc.edu

---

[1]Visiting Astronomer, Kitt Peak National Observatory, National Optical Astronomy Observatories, which is operated by the Association of Universities for Research in Astronomy, Incorporated, under cooperative agreement with the National Science Foundation

[2]Hubble Fellow

## Abstract

A refined sample of 64 variable objects with stellar image structure has been identified in SA 57 to $B \sim 22.5$, over a time baseline of 15 years, sampled at 11 distinct epochs. The photometric data typically have a root-mean-square error at $B = 22$ of only 0.05 mag. Thirty-five quasars in this field have already been spectroscopically confirmed, 34 of which are among the sample of variables. Of the other variables, 6 are known spectroscopically to be stars, 10 additional objects are stars based on reliable detection of proper motion, and 1 is spectroscopically a narrow-emission-line galaxy. Of the 13 remaining variables, it is argued that they are a mixture of distant halo subdwarfs and quasars with star-like colors. We compute the ensemble average structure function and autocorrelation function from the light curves in the respective quasar rest-frames, which are used to investigate the general dependences on apparent magnitude, absolute magnitude, and redshift.

*Subject headings:* QSO: – variability – proper motion – evolution – redshifts

## 1. INTRODUCTION

This paper is one of a series that explores the statistical properties of faint quasars in Selected Area 57. Kron and Chiu (1981) reported initial faint spectroscopy for a sample of stellar objects chosen by (lack of) proper motion. Colors, slitless spectroscopy, and variability information was also available for a larger sample, and a preliminary assessment was made concerning the efficiency of these techniques for finding faint quasars. Koo, Kron, and Cudworth (1986; KKC) later published a list of 77 stellar objects that appeared to have colors unlike stars of comparable magnitude. Their spectroscopic results confirmed that a high fraction of the objects were indeed quasars. KKC also showed that most of the quasars were variable over a span of several years, consistent with 8 out of 10 being detected to be variable by Kron and Chiu (1981). Koo and Kron (1988; KK) published additional redshifts for this same sample, and derived the redshift-dependent luminosity function, extending previous determinations to fainter luminosities at a given redshift. This showed that the well-known steep slope of the luminosity function at high luminosities breaks to a shallower slope a fainter luminosities, a result seen with better statistics by Boyle *et al.* (1988, 1990), but over a smaller range in redshift. Trevese *et al.* (1989; T89) made an independent study of the variability of faint stellar objects in SA 57 and found that the objects selected by variability overlapped strongly with the KKC color-selected list, thus placing an interesting limit on the incompleteness of the color-selection technique for finding quasars. Three new quasars that were overlooked by KKC were found by Majewski *et al.* (1991) by selecting objects based on variability and lack of proper motion, consistent with the expected incompleteness in the KKC list.[3] Kron *et al.* (1991) have devised an objective procedure for color-selection of quasar candidates (cf Warren *et al.* 1991), but this new sample differs only incidentally from KKC.

Quasar light curves show structure over a wide range of time scales and with a wide range of amplitudes (e.g. Pica and Smith 1983). The long-standing hope has been that physical constraints on the nature of the central quasar engine could be derived from arguments based on variability studies, but the great variety shown in the behavior of the well-studied cases has so far led only to rather general conclusions. Even simple phenomenological results, such as whether more luminous quasars show more variability, have apparently discordant observational bases (cf Uomoto, Wills, and Wills 1976 and Giallongo, Trevese, and Vagnetti 1991, GTV), suggesting that the details of the observational selection and the time sampling may qualitatively affect the conclusions.

[3]A fourth quasar found by Majewski *et al.* (1991) at z=3.54 was fainter than the KKC limit.

Specifically, GTV showed that a measure of variability that combines the effects of correlation and cosmological time dilation can make a positive variability – redshift relation appear to be negative.

Variability in quasar fluxes may also indirectly affect other physical interpretations. The number-magnitude and magnitude-redshift relations are biased by variability (Bahcall 1980), as is the derived luminosity function and its dependence on redshift (GTV).

In addition to studying variability in a sample of known quasars, it is feasible to study the variability properties of a magnitude-limited sample of all objects (or at least objects with stellar or nearly stellar images) in a field, and then determine what fraction of these are quasars. Such surveys have been carried out recently also by Hawkins and Véron (1993) and by Cimatti, Zamorani, and Marano (1993), who give a good review of earlier work.

In this paper, we extend the results of variability of faint quasars of T89 by i) adding new data from more recent epochs; ii) using a better threshold for variability; iii) exploring correlations of variability with color and proper motion; and iv) computing statistical measures of the variability as a function of time difference in the quasar rest-frame.

## 2. OBSERVATIONS

Prime focus IIIa-J plates of SA 57 have been obtained with the Mayall 4-m telescope at the Kitt Peak National Observatory since 1974. The digitization, object selection, image classification, and photometry have been described in Koo *et al.* (1986) and in T89. The object magnitudes are on the photographic $J$ (or $B_J$) system, where $B \sim J + 0.1$. The most significant change with respect to T89 is the addition of new photometric data that extend the time baseline to more recent epochs. Table 1 gives the list of all the plates that now contribute to the analysis. Altogether we have measured 14 plates spanning 15 years at 11 independent epochs.

T89 restricted the area that was analyzed to the overlap with the KKC area, which ensures that all objects have multi-color photometry from Koo's (1986) photometric catalogue of SA 57. They further restricted their sample to objects with stellar image structure, with the threshold in image size set to include essentially all of the objects independently classified by Koo (1986) as stars. The magnitude limit (defined by plate MPF 1053 and within a small photometric aperture, as described in T89) was chosen to be faint enough that essentially all of the KKC objects (limited by $J_K = 22.5$) were included. These criteria yielded a sample of 694 stellar objects, and the present study adopts this same sample. We have added the more recent photometric epochs to the existing database.

Spectra for candidate quasars have been obtained over a period of several years at Kitt Peak National Observatory with the Mayall 4-m telescope. Most of this spectroscopy has been included in T89. Since then, redshifts for 3 previously unconfirmed quasars in the KKC list have been obtained. The new spectroscopy has also provided additional exposure time for some of the T89 sample, which has resulted in two more quasar redshifts. In addition, as mentioned in the Introduction, 4 quasars not in the KKC list have been discovered by Majewski *et al.* (1991), 3 of which are within the magnitude limit of our current sample.

We have made a comprehensive re-evaluation of the spectra of the objects considered by T89 and we have formulated a new measure of spectroscopic quality, Q, that is based on the number of spectral features that are definitely detected and the number of features that are likely to be real (see footnote to Table 3). The quality factors are intended to provide a consistent measure of reliability for the stated redshift. It is important to note that with the exception of 3707 and 4882, all quasars with Q=2 have only one definite line identification. The line identification (and assigned redshift) in some of these cases utilized the absence of other emission lines to choose among possible sources of emission (e.g. MgII vs. CIV). As a consequence of this review and new data, the redshifts reported here (Table 3) are in a number of cases slightly different in the last decimal place. The redshifts for 6 objects in T89 have been substantially revised, but only 3 of these changes are significant in the sense that they had high quality factors (i.e. as reported in KK):

In the case of NSER 5141 (KKC 34), the spectroscopic quality assigned in KK indicated two detected features and a redshift of 2.30. However, they noted that this assignment was on the basis of spectra reported by Kron and Chiu (1981), who themselves indicated these features as uncertain. A more recent spectrum clearly reveals two features which correspond to CIII] and MgII at z=1.09.

For NSER 10028 (KKC 1), no new spectra have been obtained, but we have re-processed the existing spectra since KK. The reduction used for the KK analysis shows a definite feature at 4618Åand a probable feature near the sky line at 5577Å; these features were identified as CIV and CIII], respectively. Upon more careful reprocessing, the second feature disappeared, and we have reassigned this line to MgII on the basis of lack of emission at longer wavelengths (extending to 7500Åin the observed frame).

Finally, NSER 360 (KKC 54) appears to be a puzzling object. The first two spectra taken with the Cryogenic Camera in March 1983 show a nearly featureless, blue continuum with some evidence for H$\alpha$ absorption, and this object was assumed to be a blue subdwarf by KKC and KK. However, a spectrum taken in January 1991 (RC spectrograph) shows a strong feature near H$\beta$ in the observed frame, but shifted by $\sim$2000 km/s to the red. We have currently assigned MgII to this feature, since if it were either CIII] or CIV we should have detected the other line, and in the latter case Ly-$\alpha$ should have beed detected as well. We cannot rule out that this object is a peculiar, variable star.

The updated spectroscopic results are given in Table 3, including 6 stars and 1 narrow-emission-line compact galaxy. (Table 3 reports the spectroscopic results for the sample of variables, which includes all but one of the known quasars within our magnitude limit. This one additional QSO, NSER 8169, is appended to Table 3 for completeness.) In the following analysis, we will ignore any distinctions in the way the quasars were originally selected for spectroscopy.

The proper motions for most of the 694 stars in T89 have been measured by Majewski (1992), reckoned in a frame such that the average motion of the known quasars is defined to be zero. As discussed later, these motions can be used to discriminate between stars and quasars even when spectroscopic verification is lacking.

In summary, we have high-quality multi-color and multi-epoch photometry and astrometry for a magnitude-limited sample of faint stellar objects in SA 57. Colors, proper motions, and image structures are known for each object. Many objects have been measured spectroscopically, and so far there are 35 quasars with reliable redshifts to $B = 22.5$. The distribution of redshifts is approximately uniform over the range $0.5 < z < 3.1$, and there is thus a wide range of luminosities represented in the sample.

## 3. ANALYSIS

### 3.1. Variability Measurement

We have designated MPF 1053 to be the plate and epoch with respect to which the others are measured. This fiducial plate is one of the best plates in the whole SA 57 IIIa-J plate collection, and it coincidentally establishes the first epoch. The positions of faint objects were originally found on this plate, and then photometry was performed on the other plates at these same positions. Thus, a variable object that happened to be especially faint at the epoch of MPF 1053 may not be included in the catalogue. For simplicity, the magnitude $m_4$ and image structure parameter $m_1$ - $m_3$ are similarly taken from the measurement on MPF 1053 alone. These parameters are discussed in T89: $m_4$ is the aperture magnitude inside the 4th radial step, r $\sim$ 1.1 arcsec, and the image structure parameter $m_1$ - $m_3$ measures the degree of central concentration of the light.

Because of the non-linearity of the photographic photometry, and the differences in this non-linearity from one plate to another, we have chosen to transform the object magnitudes from each plate to the system of MPF 1053. This is done for the sample of 694 stellar images by computing $\Delta$(i) = $m(i)$ - $m(1053)$ for each plate $i$; rejecting very deviant points in the $\Delta$ versus $m_4(1053)$ plot; computing the average instrumental magnitude difference that applies to each object based on the 61 objects closest to it in magnitude; and finally correcting $m(i)$ by this smooth relation. Henceforth $m(i)$ will refer to the values of $m(i)$ corrected for this average offset. As in the T89 analysis, $m(i)$ refer to magnitudes measured in optimal apertures, which are in turn a function of $m_4$.

This procedure is not rigorous – for example, a fit in $m(i)$ - $m(1053)$ versus $m_4(i)$ (as opposed to $m_4(1053)$) would have yielded a different set of corrections – but this is unlikely to matter since we only require sensitivity to variability. Moreover, we have made no attempt to remove the non-linearity present in the MPF 1053 photometry, the effect of which can be seen as a bunching of points at the bright end in Figure 1.

Thus, our basic variability data consist of corrected $m(i)$'s for each object at each epoch, such that the mean difference in corrected magnitudes for all stars of similar magnitude on the plate with respect to MPF 1053 is zero, by construction. The plates yield photometric measures with a range of quality, as is clear from the spread of the $\Delta$'s. This spread is the predominant factor in the assignment of the weights to each plate (Table 1).

This database, with its relatively small photometric errors, relatively long time baseline, and relatively good sampling, provides new leverage on the correlation of variability with redshift and absolute magnitude (Uomoto, Wills, and Wills 1976; Bonoli *et al.* 1979; Pica and Smith 1983; Netzer and Sheffer 1983; Cristiani, Vio, and Andreani 1990).

Table 4 gives the magnitude deviations for each object on each plate that were determined to be variable as described below. Specifically, the tabulated numbers are the differences

$$\delta(i) = m(i) - <m>,$$

where as described earlier $m(i)$ are the optimum-aperture magnitudes corrected to the magnitude scale of MPF 1053, and $<m>$ is the unweighted average of the magnitudes for the 14 plates. The light curve in aperture 4 (radius 1.1 arcsec) can be recovered from

$$m_4(i) = \delta(i) + m_4(1053) - \delta(1053),$$

where the last two terms come from Table 3 (sixth column) and Table 4 (third column). Blank entries in Table 4 correspond to bad plate or scan regions, and are given zero weight.

Our variability index $\sigma *$ is the same as described in T89. It is essentially a weighted root-mean-square magnitude fluctuation, where the weights are given in Table 1. The prescription to recover the values for $\sigma*$ listed in Table 3 from the data in Table 4 is as follows. First, the three pairs of plates at a common epoch are combined to form weighted averages, so that the 14 plates are reduced to 11 epochs. Since the idea is that the plate pairs measure the same flux, we take advantage of the opportunity to filter out glitches by demanding that the quantity $\mid \Delta_1 - \Delta_2 \mid$ be smaller than 0.25, 0.30, and 0.30 mag for the plate pairs MPF 1561/2, 3919/21, and 4181/89, respectively. ($\Delta_1 = m(1561)$ - $m(1053)$, for example.) Otherwise, the plate with the largest deviation is given weight zero for that object. We then compute the weighted mean magnitude over the 11 epochs for each object, and then give weight zero to the largest single deviation from the mean greater than 0.67 mag (this is the main filter for single-plate glitches). Finally, we re-compute the mean magnitude using the revised weights, $w$, with respect to which we calculate the weighted r.m.s, which is $\sigma*$:

$$\sigma*^2 \;=\; \sum_i \,(w_i\; \delta_w(i))^2 \;/\; (\sum_i \; w_i^2),$$

where

$$\delta_w(i) = m(i) - < m >_w,$$

and $< m >_w$ is the weighted average of the magnitudes for the 11 epochs. Aside from the addition of the new plates, the only part of this procedure that differs from T89 is a slightly different $\mid \Delta_1 - \Delta_2 \mid$ threshold for the pair MPF 3919/21.

In T89, we used a fixed threshold $\sigma* > 0.1$ to define the sample of nominally variable objects. Here, we refine this by adopting a threshold that, at each magnitude, is $2\sigma$ above the mean $\sigma*$ for non-variable stars. This threshold is shown in Figure 1 and is tabulated in Table 2. There are a total of 64 variables so defined, larger than the 50 of T89, partly because of the inclusion of smaller-amplitude variables at the brighter magnitudes. The mean $\sigma*$ for the non-variable stars is our estimate of the weighted photometric random error, $\sigma_n$; since we will use this later, its value as a function of apparent magnitude is also recorded in Table 2. The 64 variables are listed in Table 3, and the measurements on the individual plates are given in Table 4. The column headings $N_{694}$, NSER, and KKC indicate serial numbers in the T89, Kron (1980), and KKC catalogues; the remaining column headings are defined throughout the paper. Colors in Table 3 are from KKC and match $J_K$.

Figure 1 apparently shows an upper envelope that increases with increasing magnitude. This is visible also in the earlier version of this plot (T89), and the effect is present both for the KKC objects and for the remainder. The envelope is not matched by scaling the magnitude-dependent random errors. Over this range of magnitudes, there is essentially no change in the quasar redshift distribution (KK), and the objects that define the upper envelope in Figure 1 for which we have redshifts show the usual wide range of redshifts and intrinsic luminosities. Hence, it is unlikely that the apparent trend can be explained by differences in the quasar population that depend on apparent magnitude, a point to which we will return later.

The 35 known quasars in the field are indicated in Figure 1, and it can be seen that the new selection threshold excludes only one of them. Note that the excluded quasar has $J(r = 1.''1) > 22$ and is close to the variability selection threshold. Hence, if one had data with somewhat smaller errors or better sampling, *all* quasars may be detectable by their variability. The 34 variable quasars form an important subset of the 64 variables since the light curves in their respective rest-frames can be computed (section 3.3).

Of the 64 variables, 24 are not listed by KKC (but 3 are beyond the $J_K = 22.5$ limit of KKC), meaning that in general these have the colors of normal stars. Six of the 24 non-KKC objects have been observed spectroscopically and have been shown to be stars (3 dM stars and 3 bluer stars); 2 of the 24 have been observed spectroscopically and shown to be quasars, and one is a compact narrow-emission-line galaxy. There remain 15 variables without spectroscopic identification that are not in the KKC list (and 8 variables without spectroscopic identification that are in the KKC list).

### 3.2. Proper Motion Measurement

To refine further the identities of the variables without spectroscopic identification, we take advantage of Majewski's (1992) proper motion measurements. Three of the 64 variables have no astrometric data because they are fainter than Majewski's selection limit. Table 3 includes entries for $\mu_0$, the proper motion in units of arcsec/century towards Galactic longitude $l = 0$; $\varepsilon_{\mu_0}$, the error in the measurement of this quantity; $\mu_{90}$, the proper motion towards $l = 90$; and $\varepsilon_{\mu_{90}}$. The quantities $\mu_0$ and $\mu_{90}$ are reckoned with respect to a reference frame such that the mean motion of the known quasars is zero. Our adopted criterion for a detection of proper motion is 4 sigma in the combined directions. Specifically we define a "proper motion index," *pmi*, where

$$pmi = [\ (\mu_0/\varepsilon_{\mu_0})^2\ +\ (\mu_{90}/\varepsilon_{\mu_{90}})^2)\ ]^{1/2},$$

and the criterion for a detection of proper motion is $pmi \geq 4.0$.

Figure 2a shows $\mu_0$ versus $\mu_{90}$ for the 35 known quasars and the 6 known stars. All of the quasars have $pmi < 4.0$, and only one has $pmi > 3.0$. Figure 2b shows the same diagram for 20 variables without spectroscopic identification that have astrometric measurements. The solid symbols correspond to objects in the KKC list (i.e., non-stellar colors), and the open symbols correspond to objects with stellar colors. In general, the objects with $(\mu_0^2 + \mu_{90}^2)^{1/2} > 1$ arcsec/century are red, and the objects in the clump at smaller motions are bluer. The centroid of the blue clump is offset to negative values of $\mu_{90}$ with respect to the quasars in Figure 2a, suggesting that some objects in the clump of Figure 2b may be distant halo stars.

The distribution of colors and motions for the 10 objects in Figure 2b with $pmi \geq 4.0$ are consistent with what is expected for a population of ordinary stars at this magnitude (Majewski 1992). Some of these stars are red and are presumably dM stars that are variable because of having active chromospheres. Cimatti, Zamorani, and Marano (1993) also found a number of variables to be probable M stars. The other moving variables without spectroscopic data are presumably similar to the two bluer stars that have good spectra: NSER 10496 and 13436 are spectroscopically apparently normal dwarf stars.

It is instructive to look at the color distribution of the objects according to proper-motion category. Figure 3 shows the $U - J$ versus $J - F$ diagram, with small circles representing the $pmi < 4.0$ objects. For the objects with $pmi \geq 4.0$, the medium-sized circles indicate objects with proper motion $\mu < 1$ arcsec/century, and the large circles indicate objects with $\mu \geq 1$ arcsec/century. Figure 3 shows that colors and proper motions are related, with the $pmi > 4.0$, $\mu < 1.0$ objects tending to be distributed around the tip of the normal subdwarf sequence.

In summary, the proper motions, and the correlation of proper motion with color, suggests that many of the spectroscopically unidentified variables are Galactic stars. The individual classifications by proper motion and by color can be checked in the cases where we do have spectroscopic identification. There remain some ambiguous classifications, namely the blue objects with small proper motion, but there is enough information to limit the total number of quasars that may still remain spectroscopically unidentified. An important case concerns quasars at very high redshift, which are expected to have red $J - F$ colors. However, no object in this part of the color-color diagram has $pmi < 4.0$ (Majewski *et al.* 1991), which limits the number of such objects that could have been missed by color or variability search techniques.

As discussed by KKC and KK, the color-selected samples of candidate quasars contain also objects of lower redshift that have narrow emission lines, consistent with the spectra of galaxies with photoionized gas. Upon more detailed study of the image structure of 7 of these objects, some extension has been found (King *et al.* 1991, Koo *et al.* 1994). These so-called compact narrow-emission-line galaxies would not be expected to be variable, and indeed there is an excellent correspondence between the KKC objects that are not variable and the sources spectroscopically identified with galaxies. Only one of the 11 such galaxies known in SA 57 appears among the 64 variables, namely NSER 11867.

That 34 out of 35 known quasars appear in our objectively defined sample suggests that high completeness can be achieved using the technique of variability searches with high-quality photometry. It is of course possible that there are other quasars in the field that are not variable and which were never targeted for spectroscopy. However, in general these would have to have normal stellar colors because otherwise they would have appeared in the list of KKC. But, the majority of the objects with normal stellar colors are observed to have $pmi > 4.0$ (Majewski *et al.* 1991). A quantitative discussion of the completeness (and reliability) of the SA57 sample was given in T89; the results reported here are similar since there is a strong overlap of the objects with known redshift.

### 3.3. Ensemble Variability Statistics

We have computed two statistical measures of variability averaged over the whole ensemble of the variables or some subset of the variables, such as the spectroscopically confirmed quasars. This procedure naturally assumes that variability in different quasars is due to the same underlying driving agent. (If the variability is due to a stationary random process, then the light curves of the separate quasars are equivalent to having separate time series for the same quasar.)

The non-linearity in the magnitude scale at bright magnitudes ($m_4 < 19.5$) has the potential to introduce apparently different variability behavior in the brighter quasars with respect to the fainter quasars. However, we expect this effect to be small, partly because only one quasar is brighter than $m_4 = 19.5$, and partly because the difference between the instrumental magnitude and the actual magnitude is only about 0.5 mag at $m_4 = 19$.

The first ensemble statistic is the first-order structure function (Bonoli *et al.* 1979; Simonetti, Cordes, and Heeschen 1985), defined here by

$$S^2(\tau) \ = < [m(t+\tau) \ - \ m(t)]^2 > \ - 2\sigma_n^2,$$

where $m(t)$ is the magnitude at epoch $t$ reduced to the magnitude scale of plate MPF 1053, $\tau$ is the time lag, and the average is taken over the ensemble of variables. This statistic was discussed by Trevese and Kron (1990) and by GTV as applied to the data in T89. The parameter $\sigma_n$ is the contribution of the photometric random noise to the structure function; its value is not critically important because according to Figure 1, $\sigma_n$ at each magnitude is well below the typical level of the variations. It acts as a zero-point offset to the amplitude of the structure function, without changing the shape of the structure function.

Figure 4a shows the apparent structure function for the known quasars.[4] The error bars indicate the root-mean-square dispersion of the data for each of the discrete time lags $\tau = t_i - t_j$. Figure 4b shows the same for the 15 variables with $pmi \geq 4.0$ (here and in the following figures we bin the data in intervals of time lag, since the higher time resolution of the unbinned data is not useful). Figure 4c shows the same plot for the 10 objects without either a detected proper motion or spectroscopic confirmation. The quasar sample (Figure 4a) shows a distinct rise in the structure function within the first three or four years, followed by a more gentle rise to the limit of the data at long time lag. As expected from the positions of the moving stellar variables in Figure 1, the amplitude of their structure function (Figure 4b) is much lower, and there is no evidence of the same kind of correlated variability over these time scales. Since Figure 4c shows a behavior that is intermediate between Figure 4b and Figure 4a, it seems reasonable to suppose that the 10 unidentified objects with $pmi < 4.0$ comprise a mix of halo stars with relatively low proper motion and a few additional quasars, consistent with the inference from the shift in the centroid of small proper motions in Figure 2b.

It is more interesting and more physically meaningful to compute the structure function in the rest-frame of the quasars. We assume that the quasars for which we have redshifts are representative of all of the quasars, at least from the point of view of variability. We use the sample of all 35 known quasars, including the one that was not detected to be variable.

Figure 5 shows the rest-frame structure function for the sample divided in different ways. In each case, the number of quasars that contribute to each bin declines for the bins at larger time lag. To ensure that a reasonable number of quasars populate each bin, we plot only those bins which have at least two-thirds of the sample that is under consideration. In the first year, we plot bins with intervals of 0.25 year in order to resolve changes on relatively short time scales; at longer time lags we use bin intervals of 0.5 year. The error bars give the r.m.s. spread of the individual points around the mean. Figure 5a shows the parent sample of 35 quasars, and Figures 5b and 5c compare the sample divided evenly by redshift, where the median redshift is 1.34. The low-redshift sample (5b) allows the maximum time difference to be sampled. On the other hand, Figure 5c, for the high-redshift half of the sample, has fewer bins but more quasars contributing to each bin. Note that the amplitude is higher for the high-redshift subsample, as expected from GTV's suggestion that $\sigma*$ correlates positively with redshift. The respective rise times for the two subsamples are similar: most of the total variability is reached after 2 years.

Figure 5d with 5e compare, again for the 35 quasars with known redshifts, the half of the sample with the smallest luminosities to the half with highest luminosities, where the median luminosity is $M_J = -23.92$ ($H_0 = 50$ km sec$^{-1}$ Mpc$^{-1}$, $q_0 = 0.5$, and $f_\nu \sim \nu^\alpha$ with $\alpha = -1$). The lower-luminosity quasars have higher variability amplitude, on average.[5]

[4]In Figures 4 and 5 we actually plot just $< [m(t+\tau) \; - \; m(t)]^2 >$.

The overall shape of the structure function in Figure 5a suggests that about half of the amplitude is achieved in less than one year in the rest-frame, but that the curve is still slowly rising to and presumably beyond 5 years time lag. A fit to Figure 5a can be made with a function of the form

$$f(\tau) = A\ (1\ -\ exp(-\tau/T))\ +\ 2\ (0.06)^2,$$

where the last term is the offset due to the measurement error $\sigma_n$, the adopted value for which is typical for the quasars. The fit gives $A \sim 0.16$ and $T \sim 1.0$ year. (Note that a similar exercise by Trevese and Kron (1990) was a fit to $S$, not $S^2$.)

We have made a rough estimate of the significance of the differences between the samples of Figures 5b and 5c, and Figures 5d and 5e, as follows. For each sub-sample, we fit the data using the same functional form as shown in Figure 5a to determine the two principal parameters $A$ and $T$, their errors ($\epsilon$), and their associated confidence intervals. The results are given in Table 5, where the number of points used refers to Figure 5. The probability levels are not exact because we assumed that $A$ and $T$ are uncorrelated, and moreover the procedure implicitly assumes that the points are statistically independent. A more sophisticated estimator could be derived but the information content of the data probably do not support the effort. Nominally, the result is that the high-redshift sample differs from the low-redshift sample in having a variability amplitude $A$ that is twice as great with similar characteristic time $T$, while the high-luminosity sample differs for the low-luminosity sample in having half the amplitude and half the characteristic time scale; the difference in $A$ is statistically more significant than the difference in $T$.

The other ensemble statistic that we will consider is the discrete autocorrelation function (Edelson and Krolik 1988). For a given object, we first compute its mean magnitude $< m >$ and the quantity $\delta(t_i)$ at each discrete epoch $i$, as described in Section 3.1. We also compute the quantity

$$\sigma^2 = \frac{1}{N}\sum_i \delta(t_i)^2.$$

---

[5]To check for possible effects of saturation on the photographic plates we conservatively have excluded all (7) objects with $m_4(1053) < 20.5$ from, for example, Figures 5d and 5e. The amplitudes of the structure function increase slightly, but are consistent with the original figures.

The correlation function is defined to be

$$C(\tau) \; = < \delta(t+\tau)\; \delta(t) > \; / \; (<\sigma^2> \; - \; <\sigma_n^2>),$$

where again the average over the ensemble of variables is indicated by the brackets. As for the structure function, when working with the rest-frame time lags, we bin the data in intervals of time lag, with extra time resolution at shorter lags.

In principle, $\sigma^2$ is computed separately for each object with respect to its mean magnitude in the time series as just described. However, if the variability of each quasar is a different realization of the same underlying process, then a mean value for $\sigma^2$ can be used. Adopting a mean value for $\sigma^2$ gives quasars with greater $\sigma*$ proportionally more weight in the discrete correlation function. We have computed the function both ways, and the results are very similar.

The binned discrete autocorrelation function is plotted in Figure 6 for the same subgroups of the quasars as in Figure 5. The correlation drops to zero at about 1 year lag, and thereafter gradually becomes more negative. This behavior can be understood considering that the magnitude differences $\delta(t)$ are defined with respect to the mean magnitude in the time series for each quasar. For instance, if the variation has been monotonic, the earliest and latest parts of the light curve have $\delta$'s of opposite sign. Thus the negative correlation at large time lags indicates that the variability is correlated on a time scale longer than our baseline.

One can also consider the connection between $S^2(\tau)$ and $C(\tau)$. Neglecting the noise term $\sigma_n$, one can easily show that

$$S^2(\tau) = 2 <\delta^2> \; [1 \; - \; C(\tau)].$$

Thus, when $C(\tau) \; < \; 0$, $S^2(\tau) \; > \; 2 <\delta^2>$ . The average level of $2 <\delta^2>$ is indicated in Figure 5a. The point where $C(\tau)$ becomes negative may not be accurate since we do not have data over very long time intervals and the assumption of having sampled a stationary random process may be wrong. Nevertheless, the discrete autocorrelation function still gives information on the characteristic time scale for variability, and it has the important property of being independent of the variability amplitude.

## 4. CONCLUSIONS

The statistical dependence of quasar variability (amplitude and time scale) on redshift, or on intrinsic parameters such as color or luminosity, can be addressed with sufficiently large and sufficiently representative samples (Bonoli *et al.* 1979, Cristiani *et al.* 1990, T89, GTV, Hook *et al.* 1991, Hawkins and Véron 1993). Cimatti, Zamorani, and Marano (1993) have recently presented a study of variability of a sample of objects with stellar images with a similar magnitude limit from similar data. Their analysis has only three epochs spanning two years, and is aimed at studying the efficiency of quasar selection under these conditions. Unlike us, they use three bands ($UJF$) jointly for identification of likely variable objects, which in their application significantly improves the sample selection. For the 45 quasars in their spectroscopically complete sample, Cimatti, Zamorani, and Marano (1993) find no statistically significant correlation of their variability index with either redshift or with luminosity.

Hook *et al.* (1991) analyzed a sample of almost 300 quasars to $B = 21$, a time baseline similar to ours. For their sample, they concluded that 1) there was only a weak anticorrelation of observed-frame variability amplitude with redshift; 2) there was a stronger anticorrelation with absolute magnitude (more luminous quasars are less variable); 3) there was no evidence for any dependence of variability amplitude on apparent magnitude. As they note, the first conclusion could be partly an effect of time dilation, and since for magnitude-limited samples redshift and absolute magnitude are correlated, the same expectation holds for the second conclusion. However, when the binned $<|\Delta_m|>$ of Hook *et al.* (1991) is plotted against *rest-frame* time interval, the amplitude nevertheless remains anticorrelated with absolute magnitude. (It is well to keep in mind that the existence of a limiting magnitude imposes a correlation: quasars binned by $M$ will have different $<z>$.)

Our sample of quasars is substantially smaller than that of Hook *et al.* , but our time sampling is finer (eleven epochs as opposed to five epochs), and our photometric errors are smaller (the difference between photometry on Schmidt plates and on 4-m plates). Most significantly, our sample is weighted to quasars with lower luminosity. They find a real difference in variability amplitude only for $M_B < -27$, yet we have only two quasars brighter than $M_B = -26$. The difference in luminosities arises because half of Hook *et al.* 's sample is brighter than about $B = 19.2$, whereas half of our sample is fainter than $B = 21.5$, and their median redshift is about 1.8, whereas ours is 1.34.

Figure 7 shows the data in a log $z$ – $M_J$ plane ($H_0 = 50$ km sec$^{-1}$ Mpc$^{-1}$, $q_0 = 0.5$, $\alpha = -1$), with the magnitude limit $J_K = 22.5$ indicated. This shows a generally uniform distribution, except for the absence of apparently bright quasars (lower left-hand corner). The median values of $M_J$ and log $z$ are also shown. We have computed the ensemble average rest-frame structure function and correlation function for the quasars in each of the four quadrants of Figure 7, which yields estimates of the long-term amplitude and time scale, respectively, for each subgroup. The result is that the amplitude for variability appears to increase most strongly from the lower left of Figure 7 to the upper right, i.e., in the direction of increasing apparent magnitude. (This result is already apparent from the increase in the median $\sigma *$ with apparent magnitude in Figure 1.) If the sample is split in halves by apparent magnitude, the amplitude of the structure function is larger by more than a factor of two for the fainter half. On the other hand, in the perpendicular direction, the amplitude for the structure function appears similar for the low-luminosity, low-redshift quadrant and the high-luminosity, high-redshift quadrant. As already discussed with respect to Figure 5bcde, comparing the low- and high-redshift halves, the ensemble average amplitude is higher at larger redshift, and comparing the low- and high-luminosity halves, the ensemble average amplitude is higher at lower luminosity.

The correlation function measures the characteristic time scale relatively cleanly because the variability amplitude is factored out. Considering as before the sample as divided into the quadrants of Figure 7, the time lag at which the ensemble average correlation function has a value of 0.5 (say) increases from the high-redshift, high-luminosity quadrant to the low-redshift, low-luminosity quadrant. Comparing the other two quadrants, the correlation functions appear to be similar (although the uncertainty is large because there are few quasars in each of these quadrants). To explore this a bit further, we defined a parameter $\xi = M - 5 \log z$, which increases from the upper left of Figure 7 to the lower right, and which is almost orthogonal to the direction of increasing apparent magnitude. When the sample is divided into halves by this parameter ($\xi <$ or $> -24.4$), as anticipated the difference in variability time scales is apparent (Figure 8ab); this result gives a somewhat different - but generally consistent - picture than that derived from Table 5. As discussed above, the amplitude of the structure function does not appear to be substantially different for these two sub-samples (Figure 8cd), confirming that the amplitude is primarily correlated with apparent magnitude.

An obvious concern with this conclusion is whether our photometric errors increase with increasing faintness in such a way that the above result is only apparent. For example, the nominal selection limit shown in Figure 1 is based on Gaussian errors, but the shape of the error distribution, and the wings in particular, could in principle change with apparent magnitude. However, unexpectedly large excursions in the measurements of faint-object magnitudes would also have the effect that more stars would be erroneously included above the selection threshold. If anything, it appears from Figure 1 that the opposite is the case. Therefore, we have no reason to doubt that we are measuring a real increase in variability amplitude with increasing apparent magnitude. We stress the importance of data with small random photometric errors (see also the discussion in Cimatti, Zamorani, and Marano 1993). It appears that quasars typically vary with an amplitude of $\sim 0.3$ mag, and the results of a survey – even qualitatively – are likely to be critically dependent on the size of the random photometric errors compared to this value.

A consistent picture for the behavior of the variability amplitude is that there are two dependences: quasars are generally more variable at lower luminosities, and they are generally more variable at higher redshifts. This can be understood on the basis of the following elementary model. Suppose quasars consist of a number of variable emitting regions or sub-units, each with similar average luminosity. The total luminosity would then be expected to have smaller variation $\Delta L/L$ in more luminous quasars because there are more independent sub-units. The dependence on redshift would occur if quasars were more variable at shorter rest wavelengths, since with a fixed observing band high-redshift objects are sampled at higher rest-frame frequencies. This hypothesis has some empirical basis (see e.g. discussion in GTV). Moreover, it would not be surprising if quasars at higher redshift were found to be more variable because of the expected higher space density of the surrounding gas clouds, or perhaps because of the relative youth of the central engines. In any case, the conjecture is that these two dependences reinforce to produce a strong dependence of variability amplitude on apparent magnitude.

It may also be possible to understand the dependence of the characteristic time scale for variability on luminosity in a similarly elementary way. If the sub-units interact with each other or with the general radiation field, and if the interaction rate is higher in the more luminous quasars, the interaction might tend to limit the maximum coherence time for an individual sub-unit to vary. Thus, we might expect the net correlation time for quasar variability to be greater for lower-luminosity quasars, as apparently observed.

This work was supported by NSF AST-8814251. RGK gratefully acknowledges support from the Italian CNR. Assistance has been provided by John Smetanka. SRM and MAB acknowledge support from NASA through grant numbers HF-1036.01-92A and HF-1028.01-92A, respectively, from the Space Telescope Science Institute, which is operated by the Association of Universities for Research in Astronomy, Incorporated, under contract NAS5-26555. MAB also acknowledges support from NASA through Graduate Fellowship NGT-50677.

---

TABLE 1

PLATE JOURNAL OF SA57

| MPF | UT date | exposure (min) | HA (end) | weight |
|---|---|---|---|---|
| 1053 | 1974 May 21 | 45 | 0W 33 | 1.0 |
| 1561 | 1975 Apr 04 | 45 | 1E 13 | 0.8 |
| 1562 | 1975 Apr 04 | 45 | 0E 21 | 0.6 |
| 2176 | 1976 Dec 01 | 45 | 3E 17 | 1.0 |
| 3313 | 1980 May 17 | 20 | 1E 02 | 0.6 |
| 3622 | 1982 Jan 31 | 55 | 0W 42 | 1.0 |
| 3919 | 1984 Apr 05 | 50 | 1E 57 | 0.6 |
| 3921 | 1984 Apr 05 | 70 | 2W 01 | 0.8 |
| 3977 | 1985 Apr 25 | 60 | 3W 12 | 0.8 |
| 4119 | 1987 Jan 27 | 60 | 0E 15 | 0.9 |
| 4160 | 1987 Jun 26 | 55 | 3W 16 | 0.6 |
| 4173 | 1988 Apr 20 | 60 | 3W 08 | 0.4 |
| 4181 | 1989 Jan 10 | 70 | 0E 18 | 1.0 |
| 4189 | 1989 Jan 11 | 65 | 0E 51 | 0.8 |

TABLE 2

VARIABILITY THRESHOLD

| $m_4$ | $\bar{\sigma^*}$ | $\bar{\sigma^*}+2\sigma_{\sigma^*}$ |
|---|---|---|
| 19.0 | 0.028 | 0.058 |
| 19.5 | 0.029 | 0.059 |
| 20.0 | 0.030 | 0.060 |
| 20.5 | 0.033 | 0.063 |
| 21.0 | 0.036 | 0.066 |
| 21.5 | 0.040 | 0.080 |
| 22.0 | 0.050 | 0.100 |
| 22.5 | 0.075 | 0.145 |
| 23.0 | 0.120 | 0.210 |

TABLE 3

VARIABLE OBJECTS

| $N_{694}$ | NSER | KKC | R.A. (1950) | Dec. | $m_4$ | $J_K$ | U-J | J-F | F-N | z | Q[a] | $\sigma^*$ | $\mu_0$ | $\varepsilon_{\mu_0}$ | $\mu_{90}$ | $\varepsilon_{\mu_{90}}$ | *pmi* |
|---|---|---|---|---|---|---|---|---|---|---|---|---|---|---|---|---|---|
| 210 | 6506 | ··· | 13:05:06.94 | 29:33:35.4 | 19.13 | 18.29 | 0.49 | 1.11 | 1.13 | ··· | ··· | 0.060 | -0.913 | 0.079 | -0.133 | 0.080 | 11.676 |
| 446 | 13436[b] | ··· | 13:05:43.70 | 29:43:48.8 | 19.20 | 17.84 | -0.54 | 0.69 | 0.99 | 0.00 | 6 | 0.069 | 1.834 | 0.066 | -0.181 | 0.061 | 27.946 |
| 4 | 360 | 54 | 13:06:36.23 | 29:21:55.8 | 19.43 | 18.95 | -1.08 | 0.58 | 0.47 | 0.75 | 2 | 0.070 | 0.057 | 0.078 | 0.006 | 0.085 | 0.734 |
| 354 | 10496[b] | ··· | 13:06:07.14 | 29:39:29.2 | 19.50 | 19.08 | 0.06 | 0.84 | 0.71 | 0.00 | 6 | 0.086 | 0.617 | 0.050 | -1.074 | 0.052 | 24.059 |
| 255 | 7624 | 28 | 13:05:53.15 | 29:35:15.9 | 19.59 | 19.41 | -1.25 | 0.34 | 0.87 | 1.74 | 3 | 0.131 | -0.041 | 0.076 | 0.228 | 0.079 | 2.936 |
| 389 | 11610 | 22 | 13:05:48.99 | 29:41:10.6 | 19.81 | 19.53 | 0.57 | 0.51 | 0.45 | 3.02 | 6 | 0.175 | 0.227 | 0.106 | -0.073 | 0.094 | 2.278 |
| 47 | 1482 | ··· | 13:06:00.48 | 29:24:54.1 | 19.83 | 19.43 | -0.31 | 0.66 | 0.61 | 0.00 | 0 | 0.061 | 0.204 | 0.066 | -0.500 | 0.072 | 7.601 |
| 613 | 17750 | 19 | 13:05:45.60 | 29:50:46.7 | 19.87 | 19.75 | -1.18 | 0.52 | 0.58 | 1.18 | 5 | 0.061 | -0.149 | 0.072 | 0.079 | 0.079 | 2.298 |
| 442 | 13371[b] | ··· | 13:04:59.39 | 29:43:39.0 | 19.99 | 19.68 | 1.47 | 1.78 | 1.82 | 0.00 | 3 | 0.086 | -0.890 | 0.076 | 1.035 | 0.074 | 18.242 |
| 563 | 16713 | 10 | 13:05:26.41 | 29:49:01.7 | 20.14 | 19.69 | -0.52 | 0.48 | -0.09 | 0.99 | 3 | 0.215 | -0.034 | 0.085 | 0.058 | 0.088 | 0.771 |
| 96 | 3112 | ··· | 13:07:17.97 | 29:28:06.5 | 20.14 | 19.93 | 0.88 | 1.25 | 0.75 | ··· | ··· | 0.062 | -0.395 | 0.070 | -1.127 | 0.073 | 16.437 |
| 109 | 3544 | 50 | 13:06:26.74 | 29:28:42.3 | 20.20 | 20.14 | -1.03 | 0.22 | 0.74 | 1.81 | 3 | 0.236 | -0.150 | 0.083 | 0.224 | 0.077 | 3.425 |
| 654 | 18749 | 55 | 13:06:41.23 | 29:52:33.3 | 20.41 | 20.23 | -0.38 | 0.46 | 0.15 | 1.09 | 2 | 0.127 | -0.109 | 0.073 | 0.004 | 0.071 | 1.494 |
| 443 | 13365 | ··· | 13:07:28.58 | 29:43:36.3 | 20.45 | 20.30 | 0.37 | 1.04 | 0.62 | ··· | ··· | 0.089 | 0.574 | 0.070 | -2.195 | 0.069 | 32.851 |
| 179 | 5607[b] | ··· | 13:04:58.37 | 29:32:10.7 | 20.48 | 20.28 | -0.15 | 0.53 | 0.53 | 0.00 | 4 | 0.063 | 0.194 | 0.072 | -0.056 | 0.072 | 2.804 |
| 254 | 7567 | 37 | 13:06:09.20 | 29:35:09.7 | 20.60 | 20.48 | -0.34 | 0.13 | 0.59 | 1.81 | 3 | 0.136 | -0.113 | 0.070 | -0.034 | 0.078 | 1.672 |
| 160 | 5141 | 34 | 13:06:04.31 | 29:31:23.1 | 20.62 | 20.57 | -0.87 | 0.40 | 0.42 | 1.09 | 4 | 0.212 | -0.069 | 0.113 | 0.264 | 0.101 | 2.684 |
| 187 | 5767 | ··· | 13:05:15.05 | 29:32:25.0 | 20.73 | 20.57 | 0.05 | 0.85 | 0.77 | 0.71 | 2 | 0.166 | 0.032 | 0.092 | 0.148 | 0.082 | 1.838 |
| 172 | 5422 | 64 | 13:06:49.05 | 29:31:49.7 | 20.76 | 20.58 | -0.96 | 0.59 | 2.37 | 1.08 | 3 | 0.351 | -0.153 | 0.090 | -0.084 | 0.081 | 1.991 |
| 335 | 9980 | 25 | 13:05:50.30 | 29:38:42.1 | 20.78 | 20.20 | -0.43 | 0.14 | 0.17 | 1.54 | 3 | 0.146 | -0.043 | 0.087 | 0.051 | 0.081 | 0.800 |
| 206 | 6427 | 3 | 13:04:55.45 | 29:33:29.6 | 20.80 | 20.64 | -0.12 | 0.35 | 0.21 | ··· | 0 | 0.073 | 0.346 | 0.093 | -0.136 | 0.089 | 4.022 |
| 291 | 8688[b] | ··· | 13:04:52.58 | 29:36:51.3 | 20.83 | 20.65 | 1.17 | 1.77 | 1.82 | 0.00 | 2 | 0.130 | -0.691 | 0.083 | -1.335 | 0.078 | 19.033 |
| 37 | 1230 | ··· | 13:07:04.25 | 29:24:18.8 | 20.87 | 20.73 | -0.09 | 0.71 | 0.44 | ··· | ··· | 0.070 | 0.066 | 0.094 | -0.698 | 0.088 | 7.963 |
| 194 | 5938 | ··· | 13:05:34.55 | 29:32:39.3 | 20.88 | 20.74 | 1.46 | 1.91 | 2.31 | ··· | ··· | 0.075 | 1.575 | 0.075 | -2.114 | 0.078 | 34.286 |
| 398 | 11867[c] | ··· | 13:04:51.00 | 29:41:29.1 | 20.93 | 20.68 | -0.14 | 0.73 | 0.54 | 0.16 | 3 | 0.080 | 0.009 | 0.121 | 0.002 | 0.106 | 0.077 |
| 479 | 14544 | ··· | 13:07:29.14 | 29:45:27.7 | 20.95 | 20.82 | 0.98 | 1.48 | 0.90 | ··· | ··· | 0.082 | 1.007 | 0.091 | -0.943 | 0.083 | 15.860 |
| 157 | 5059[b] | ··· | 13:05:56.70 | 29:31:15.6 | 20.95 | 20.84 | 1.74 | 1.83 | 2.05 | 0.00 | 6 | 0.069 | -1.846 | 0.066 | -0.689 | 0.067 | 29.800 |
| 30 | 1012 | 32 | 13:06:02.22 | 29:23:44.1 | 21.01 | 20.52 | 0.43 | 0.27 | -0.42 | 2.28 | 4 | 0.277 | 0.035 | 0.224 | 0.378 | 0.174 | 2.178 |
| 459 | 13966 | 46 | 13:06:21.32 | 29:44:37.6 | 21.09 | 21.27 | -1.19 | 0.30 | 0.69 | 0.95 | 2 | 0.169 | 0.036 | 0.076 | -0.003 | 0.084 | 0.475 |
| 42 | 1339 | 44 | 13:06:17.75 | 29:24:33.3 | 21.15 | 21.35 | -1.01 | 0.45 | 0.85 | 0.75 | 2 | 0.244 | -0.104 | 0.096 | -0.105 | 0.096 | 1.539 |
| 240 | 7326 | 43 | 13:06:16.89 | 29:34:48.1 | 21.17 | 21.00 | -1.00 | 0.44 | 0.52 | 1.32 | 4 | 0.118 | -0.160 | 0.106 | 0.084 | 0.096 | 1.745 |
| 40 | 1304 | 11 | 13:05:27.05 | 29:24:30.0 | 21.21 | 21.00 | -0.19 | 0.11 | 0.45 | 2.27 | 6 | 0.204 | 0.100 | 0.108 | -0.106 | 0.110 | 1.336 |
| 371 | 11178 | ··· | 13:07:23.27 | 29:40:29.6 | 21.32 | 21.14 | -0.40 | 0.44 | 0.61 | 1.61 | 4 | 0.263 | -0.062 | 0.121 | -0.187 | 0.123 | 1.604 |
| 180 | 5643 | 27 | 13:05:52.78 | 29:32:10.4 | 21.47 | 21.13 | -0.46 | 0.39 | 0.43 | 0.98 | 2 | 0.287 | 0.010 | 0.095 | -0.014 | 0.092 | 0.185 |
| 495 | 15180 | 31 | 13:06:01.51 | 29:46:34.2 | 21.53 | 21.09 | -1.14 | 0.38 | 0.78 | 0.92 | 2 | 0.297 | -0.042 | 0.078 | -0.075 | 0.083 | 1.052 |
| 153 | 4855 | 67 | 13:06:57.73 | 29:30:55.4 | 21.55 | 22.32 | -1.40 | 0.60 | 0.79 | 1.30 | 2 | 0.394 | -0.094 | 0.115 | -0.366 | 0.127 | 2.996 |
| 627 | 18122 | 63 | 13:06:47.33 | 29:51:21.6 | 21.56 | 21.71 | -1.30 | 0.19 | 0.90 | 2.03 | 1 | 0.331 | -0.058 | 0.158 | 0.160 | 0.158 | 1.077 |
| 22 | 787 | ··· | 13:05:32.53 | 29:23:13.6 | 21.69 | 21.56 | -0.23 | 0.64 | 0.48 | ··· | ··· | 0.086 | 0.225 | 0.056 | -0.356 | 0.058 | 7.336 |
| 163 | 5185 | ··· | 13:05:55.45 | 29:31:28.2 | 21.82 | 21.62 | 0.16 | 0.57 | 0.79 | ··· | 0 | 0.213 | 0.249 | 0.167 | 0.143 | 0.165 | 1.725 |
| 438 | 13155 | 20 | 13:05:45.80 | 29:43:24.8 | 21.84 | 21.67 | -1.18 | 0.61 | 0.31 | ··· | 0 | 0.336 | 0.031 | 0.170 | -0.075 | 0.162 | 0.498 |
| 683 | 19387 | 51 | 13:06:28.60 | 29:53:56.0 | 21.86 | 21.85 | -1.30 | 0.42 | 0.46 | 1.46 | 3 | 0.291 | -0.205 | 0.180 | 0.198 | 0.175 | 1.605 |
| 114 | 3707 | 70 | 13:07:04.00 | 29:29:03.2 | 21.88 | 22.00 | -0.41 | 0.19 | 0.27 | 0.95 | 2 | 0.193 | 0.385 | 0.160 | -0.227 | 0.160 | 2.793 |
| 152 | 4882 | 23 | 13:05:49.22 | 29:30:57.1 | 21.92 | 21.86 | -0.90 | 0.55 | 1.28 | 1.47 | 2 | 0.243 | -0.011 | 0.114 | 0.086 | 0.119 | 0.729 |
| 91 | 2888 | 68 | 13:07:03.68 | 29:27:43.9 | 21.96 | 21.95 | -1.40 | 0.40 | 0.38 | 1.33 | 5 | 0.260 | 0.296 | 0.159 | -0.140 | 0.140 | 2.113 |

TABLE 3 – *Continued*

Variable Objects

| $N_{694}$ | NSER | KKC | R.A. (1950) | Dec. | $m_4$ | $J_K$ | U-J | J-F | F-N | z | $Q^a$ | $\sigma^*$ | $\mu_0$ | $\varepsilon_{\mu_0}$ | $\mu_{90}$ | $\varepsilon_{\mu_{90}}$ | *pmi* |
|---|---|---|---|---|---|---|---|---|---|---|---|---|---|---|---|---|---|
| 274 | 8169[d] | ··· | 13:05:38.70 | 29:36:07.4 | 22.04 | 21.92 | 0.00 | 0.45 | 0.29 | 0.74 | 2 | 0.091 | -0.064 | 0.139 | 0.090 | 0.167 | 0.709 |
| 414 | 12399 | ··· | 13:07:29.00 | 29:42:13.6 | 22.05 | 21.89 | -0.17 | 1.18 | 0.75 | ··· | ··· | 0.164 | -0.034 | 0.165 | -0.121 | 0.163 | 0.770 |
| 514 | 15465 | 66 | 13:06:54.35 | 29:47:00.7 | 22.12 | 21.85 | -0.71 | 0.69 | 0.82 | ··· | 0 | 0.157 | 0.147 | 0.163 | -0.076 | 0.144 | 1.045 |
| 44 | 1392 | 35 | 13:06:07.48 | 29:24:41.5 | 22.16 | 22.02 | 1.25 | 0.55 | 0.79 | 3.08 | 6 | 0.155 | 0.158 | 0.145 | 0.070 | 0.136 | 1.205 |
| 261 | 7822 | 9 | 13:05:19.62 | 29:35:34.9 | 22.17 | 22.17 | 0.08 | 0.16 | -0.18 | 2.46 | 4 | 0.401 | -0.104 | 0.130 | -0.189 | 0.132 | 1.640 |
| 384 | 11450 | 36 | 13:06:07.54 | 29:40:59.5 | 22.22 | 22.28 | -0.67 | 0.20 | -0.85 | 0.96 | 2 | 0.186 | -0.013 | 0.174 | -0.002 | 0.177 | 0.076 |
| 329 | 9877 | 30 | 13:06:00.43 | 29:38:32.0 | 22.25 | 22.02 | -0.84 | 0.51 | 0.53 | 2.12 | 4 | 0.176 | 0.088 | 0.161 | -0.055 | 0.140 | 0.673 |
| 445 | 13412 | 58 | 13:06:43.56 | 29:43:45.1 | 22.26 | 22.07 | -0.77 | 0.22 | 0.77 | 2.08 | 4 | 0.141 | -0.157 | 0.209 | -0.502 | 0.221 | 2.392 |
| 405 | 12151 | ··· | 13:04:49.91 | 29:41:52.8 | 22.32 | 21.98 | -0.04 | 0.60 | -0.17 | ··· | ··· | 0.416 | -0.284 | 0.299 | 0.238 | 0.248 | 1.350 |
| 19 | 710 | 62 | 13:06:46.69 | 29:22:56.5 | 22.35 | 22.21 | -0.29 | 0.30 | 0.51 | ··· | ··· | 0.305 | 0.205 | 0.212 | -0.539 | 0.192 | 2.969 |
| 59 | 1894 | 57 | 13:06:43.01 | 29:25:44.0 | 22.36 | 22.33 | -0.67 | 0.38 | 1.20 | ··· | 0 | 0.374 | -0.143 | 0.260 | -0.158 | 0.244 | 0.850 |
| 208 | 6442 | 56 | 13:06:41.93 | 29:33:26.8 | 22.39 | 22.25 | -1.87 | -0.38 | 1.96 | 2.12 | 4 | 0.437 | -0.312 | 0.196 | -0.316 | 0.178 | 2.384 |
| 497 | 15248 | 45 | 13:06:19.33 | 29:46:40.8 | 22.43 | 22.25 | -0.97 | 0.45 | 0.60 | ··· | 0 | 0.169 | -0.101 | 0.255 | -0.439 | 0.238 | 1.887 |
| 333 | 9934 | 16 | 13:05:43.25 | 29:38:36.9 | 22.43 | 22.23 | -0.57 | 0.40 | 0.94 | 2.53 | 4 | 0.330 | 0.170 | 0.178 | -0.347 | 0.176 | 2.191 |
| 295 | 8766 | ··· | 13:04:50.28 | 29:36:58.0 | 22.49 | 22.34 | 0.95 | 1.72 | 1.64 | ··· | ··· | 0.248 | -1.294 | 0.178 | -0.781 | 0.161 | 8.740 |
| 338 | 10028 | 1 | 13:04:49.59 | 29:38:45.8 | 22.51 | 22.29 | -0.83 | 0.16 | 0.66 | 0.65 | 2 | 0.123 | -0.302 | 0.182 | 0.325 | 0.148 | 2.752 |
| 664 | 18996 | 61 | 13:06:46.53 | 29:52:59.1 | 22.55 | 22.46 | -0.74 | 0.22 | 1.37 | ··· | 0 | 0.225 | 0.610 | 0.295 | 0.122 | 0.257 | 2.122 |
| 638 | 18424 | ··· | 13:07:09.31 | 29:51:52.9 | 22.69 | 22.77 | 0.09 | 0.93 | 0.36 | ··· | ··· | 0.276 | 0.000 | 0.000 | 0.000 | 0.000 | 0.000 |
| 258 | 7756 | 72 | 13:07:07.17 | 29:35:27.4 | 22.71 | 22.29 | -0.34 | 0.20 | 0.66 | ··· | 0 | 0.225 | 0.262 | 0.250 | -0.310 | 0.232 | 1.698 |
| 667 | 19040 | ··· | 13:06:38.90 | 29:53:07.3 | 22.77 | 22.67 | -1.17 | 1.12 | 1.13 | ··· | ··· | 0.390 | 0.000 | 0.000 | 0.000 | 0.000 | 0.000 |
| 598 | 17475 | ··· | 13:05:30.20 | 29:50:14.8 | 22.78 | 22.51 | 0.92 | 0.10 | 0.09 | ··· | ··· | 0.498 | 0.000 | 0.000 | 0.000 | 0.000 | 0.000 |

Notes –

[a] The spectroscopic redshift quality Q = min(6, $N_{probable}$ + 2 $N_{definite}$), where $N_{probable}$ and $N_{definite}$ are the number of probable and definite spectral features, respectively. Q is defined only for objects that have been observed spectroscopically.

[b] Confirmed star

[c] Compact, narrow-emission-line galaxy (see text)

[d] This object is below the variability threshold

TABLE 4

Variability Data

| $N_{694}$ | NSER | $\delta(1)$ | $\delta(2)$ | $\delta(3)$ | $\delta(4)$ | $\delta(5)$ | $\delta(6)$ | $\delta(7)$ | $\delta(8)$ | $\delta(9)$ | $\delta(10)$ | $\delta(11)$ | $\delta(12)$ | $\delta(13)$ | $\delta(14)$ |
|---|---|---|---|---|---|---|---|---|---|---|---|---|---|---|---|
| 210 | 6506 | 0.005 | -0.048 | -0.022 | -0.038 | -0.003 | -0.013 | -0.039 | -0.053 | 0.022 | -0.014 | 0.012 | -0.028 | -0.001 | 0.228 |
| 446 | 13436 | 0.057 | 0.072 | -0.052 | 0.036 | -0.012 | 0.052 | -0.083 | 0.061 | 0.033 | -0.203 | 0.034 | -0.030 | 0.079 | -0.041 |
| 4 | 360 | -0.077 | -0.077 | 0.011 | 0.099 | -0.103 | -0.071 | ··· | -0.011 | ··· | -0.047 | 0.075 | 0.135 | -0.017 | 0.082 |
| 354 | 10496 | 0.035 | 0.089 | 0.196 | 0.079 | 0.060 | -0.093 | 0.043 | -0.068 | 0.050 | -0.112 | -0.088 | -0.020 | -0.045 | -0.124 |
| 255 | 7624 | -0.088 | 0.087 | 0.268 | -0.044 | -0.118 | -0.115 | -0.106 | -0.151 | -0.020 | -0.064 | 0.073 | -0.058 | 0.160 | 0.176 |
| 389 | 11610 | 0.155 | 0.188 | 0.235 | -0.088 | 0.141 | -0.033 | -0.302 | -0.263 | -0.342 | 0.030 | 0.005 | 0.045 | 0.151 | 0.071 |
| 47 | 1482 | -0.022 | 0.031 | 0.241 | -0.038 | -0.023 | -0.105 | 0.089 | -0.016 | -0.090 | -0.012 | 0.019 | -0.073 | -0.039 | 0.046 |
| 613 | 17750 | -0.035 | 0.054 | -0.008 | -0.110 | -0.147 | 0.002 | -0.002 | 0.098 | -0.044 | 0.049 | 0.039 | 0.013 | 0.023 | 0.067 |
| 442 | 13371 | 0.050 | 0.000 | -0.025 | 0.022 | 0.033 | -0.035 | 0.012 | 0.158 | -0.013 | -0.245 | 0.065 | -0.047 | 0.088 | -0.058 |
| 563 | 16713 | 0.231 | 0.059 | -0.118 | 0.142 | 0.386 | 0.367 | -0.055 | -0.013 | 0.000 | -0.166 | -0.012 | -0.390 | -0.088 | -0.339 |
| 96 | 3112 | -0.055 | -0.037 | -0.069 | -0.007 | -0.009 | -0.096 | ··· | -0.006 | ··· | 0.045 | 0.022 | 0.131 | -0.014 | 0.101 |
| 109 | 3544 | -0.381 | -0.205 | -0.146 | -0.038 | -0.368 | -0.250 | 0.353 | 0.151 | 0.294 | 0.318 | 0.203 | -0.012 | -0.019 | 0.105 |
| 654 | 18749 | -0.087 | -0.147 | -0.234 | -0.039 | 0.117 | 0.065 | -0.117 | -0.090 | -0.126 | -0.041 | 0.263 | 0.540 | -0.058 | -0.043 |
| 443 | 13365 | -0.040 | -0.031 | -0.129 | -0.029 | 0.003 | -0.058 | -0.001 | -0.001 | ··· | -0.031 | 0.380 | 0.022 | -0.077 | 0.000 |
| 179 | 5607 | -0.012 | -0.069 | 0.046 | -0.088 | -0.035 | -0.041 | 0.271 | 0.010 | -0.034 | 0.015 | 0.113 | -0.174 | 0.025 | -0.027 |
| 254 | 7567 | -0.234 | -0.174 | 0.029 | -0.146 | 0.232 | -0.025 | 0.122 | 0.143 | 0.217 | -0.111 | -0.128 | 0.001 | 0.033 | 0.041 |
| 160 | 5141 | -0.365 | -0.130 | 0.057 | -0.205 | -0.285 | 0.022 | 0.057 | -0.044 | -0.258 | 0.038 | 0.313 | 0.434 | 0.202 | 0.168 |
| 187 | 5767 | -0.101 | -0.168 | -0.069 | -0.081 | 0.196 | 0.047 | 0.107 | 0.116 | 0.143 | 0.209 | 0.124 | -0.079 | -0.196 | -0.254 |
| 172 | 5422 | 0.484 | 0.420 | 0.534 | 0.366 | 0.238 | 0.158 | -0.271 | -0.300 | -0.001 | -0.320 | -0.371 | -0.336 | -0.301 | -0.294 |
| 335 | 9980 | 0.016 | -0.300 | -0.227 | 0.177 | -0.110 | 0.005 | 0.057 | 0.127 | 0.165 | 0.055 | 0.194 | 0.075 | -0.125 | -0.110 |
| 206 | 6427 | -0.028 | -0.075 | -0.037 | -0.003 | -0.062 | -0.013 | 0.239 | 0.002 | -0.045 | 0.029 | 0.236 | -0.240 | 0.031 | -0.032 |
| 291 | 8688 | -0.004 | -0.088 | -0.041 | 0.225 | -0.011 | -0.062 | 0.009 | -0.036 | -0.097 | -0.034 | 0.457 | -0.263 | 0.010 | -0.065 |
| 37 | 1230 | -0.029 | -0.067 | -0.023 | 0.003 | -0.031 | 0.012 | ··· | -0.031 | ··· | 0.086 | 0.009 | -0.065 | 0.030 | 0.102 |
| 194 | 5938 | 0.025 | 0.069 | 0.310 | 0.014 | -0.006 | 0.007 | 0.032 | -0.070 | -0.010 | -0.007 | -0.210 | -0.158 | 0.044 | -0.044 |
| 398 | 11867 | -0.055 | -0.130 | -0.076 | 0.004 | -0.031 | -0.030 | 0.051 | 0.090 | -0.113 | 0.084 | 0.234 | -0.065 | 0.085 | -0.052 |
| 157 | 5059 | -0.028 | 0.158 | 0.444 | 0.003 | -0.220 | -0.025 | 0.015 | -0.021 | -0.032 | -0.023 | -0.063 | -0.180 | 0.012 | -0.039 |
| 479 | 14544 | -0.048 | -0.036 | -0.095 | -0.042 | -0.020 | -0.005 | -0.001 | -0.017 | ··· | ··· | 0.311 | 0.069 | -0.091 | -0.018 |
| 30 | 1012 | -0.030 | -0.249 | 0.250 | -0.219 | 0.573 | -0.367 | ··· | -0.506 | -0.322 | 0.177 | 0.410 | 0.206 | 0.039 | 0.040 |
| 459 | 13966 | -0.040 | 0.218 | 0.152 | 0.151 | 0.051 | -0.273 | 0.016 | 0.044 | 0.170 | -0.016 | -0.317 | -0.402 | 0.121 | 0.128 |
| 42 | 1339 | -0.172 | 0.184 | 0.112 | 0.031 | -0.179 | -0.360 | -0.024 | -0.145 | -0.108 | -0.119 | 0.041 | 0.074 | 0.365 | 0.301 |
| 240 | 7326 | 0.109 | 0.021 | 0.070 | 0.058 | -0.021 | -0.155 | 0.008 | 0.051 | -0.212 | -0.154 | -0.103 | 0.111 | 0.082 | 0.129 |
| 40 | 1304 | 0.116 | 0.047 | 0.016 | -0.146 | 0.336 | -0.261 | ··· | 0.107 | 0.294 | 0.021 | 0.113 | -0.393 | -0.124 | -0.127 |
| 371 | 11178 | -0.200 | -0.321 | -0.357 | -0.191 | -0.039 | -0.229 | 0.217 | -0.040 | ··· | 0.573 | 0.400 | 0.253 | -0.146 | 0.076 |
| 180 | 5643 | -0.178 | -0.403 | -0.068 | -0.236 | -0.485 | -0.190 | -0.194 | -0.269 | -0.135 | 0.437 | 0.750 | 0.453 | 0.292 | 0.233 |
| 495 | 15180 | 0.640 | 0.409 | 0.319 | 0.106 | -0.156 | -0.071 | -0.116 | 0.013 | -0.069 | -0.441 | -0.030 | -0.372 | -0.183 | -0.044 |
| 153 | 4855 | -0.534 | 0.439 | 0.276 | 0.385 | -0.015 | 0.078 | 0.678 | 0.451 | -0.340 | 0.007 | -0.202 | -0.458 | -0.424 | -0.341 |
| 627 | 18122 | -0.433 | -0.072 | -0.190 | 0.318 | -0.605 | 0.118 | -0.508 | -0.422 | 0.126 | 0.378 | 0.377 | 0.515 | 0.176 | 0.227 |
| 22 | 787 | -0.030 | -0.006 | 0.028 | -0.003 | -0.053 | -0.041 | ··· | 0.133 | -0.086 | 0.079 | 0.237 | -0.176 | -0.054 | -0.025 |
| 163 | 5185 | -0.159 | -0.218 | -0.182 | -0.177 | 0.420 | 0.258 | 0.169 | -0.032 | -0.309 | -0.204 | 0.100 | 0.003 | 0.191 | 0.142 |
| 438 | 13155 | -0.081 | -0.050 | -0.082 | -0.291 | -0.599 | -0.590 | 0.181 | 0.183 | 0.098 | 0.352 | 0.251 | -0.009 | 0.305 | 0.336 |
| 683 | 19387 | -0.227 | -0.075 | -0.135 | 0.012 | -0.540 | -0.249 | 0.076 | 0.124 | -0.290 | -0.171 | 0.310 | 0.422 | 0.381 | 0.361 |
| 114 | 3707 | -0.317 | -0.021 | 0.018 | 0.085 | 0.428 | 0.065 | -0.066 | -0.171 | ··· | -0.364 | 0.091 | 0.352 | -0.049 | -0.051 |
| 152 | 4882 | 0.253 | 0.371 | 0.427 | 0.081 | -0.097 | -0.127 | 0.094 | -0.006 | 0.003 | -0.391 | -0.611 | -0.162 | 0.064 | 0.102 |
| 91 | 2888 | 0.080 | 0.316 | 0.312 | 0.424 | -0.244 | 0.072 | 0.036 | -0.154 | ··· | 0.096 | 0.024 | -0.411 | -0.328 | -0.226 |
| 274 | 8169 | -0.006 | 0.021 | 0.018 | 0.047 | 0.139 | 0.195 | -0.024 | 0.029 | -0.127 | -0.064 | 0.040 | -0.311 | 0.005 | 0.035 |
| 414 | 12399 | -0.084 | -0.056 | -0.087 | -0.128 | 0.098 | -0.059 | -0.058 | -0.158 | ··· | ··· | 0.638 | 0.040 | -0.152 | 0.008 |

TABLE 4 – *Continued*

VARIABILITY DATA

| $N_{694}$ | NSER | $\delta(1)$ | $\delta(2)$ | $\delta(3)$ | $\delta(4)$ | $\delta(5)$ | $\delta(6)$ | $\delta(7)$ | $\delta(8)$ | $\delta(9)$ | $\delta(10)$ | $\delta(11)$ | $\delta(12)$ | $\delta(13)$ | $\delta(14)$ |
|---|---|---|---|---|---|---|---|---|---|---|---|---|---|---|---|
| 514 | 15465 | 0.285 | 0.137 | 0.047 | -0.038 | -0.085 | -0.026 | -0.065 | 0.044 | 0.115 | -0.040 | -0.054 | 0.081 | -0.252 | -0.144 |
| 44 | 1392 | -0.041 | -0.008 | -0.063 | -0.042 | -0.079 | -0.228 | 0.281 | -0.035 | -0.031 | -0.034 | 0.183 | -0.136 | 0.171 | 0.066 |
| 261 | 7822 | 0.245 | 0.487 | 0.413 | 0.488 | 0.835 | -0.063 | -0.115 | -0.149 | -0.103 | -0.204 | -0.233 | -0.540 | -0.595 | -0.466 |
| 384 | 11450 | -0.056 | 0.227 | 0.285 | -0.004 | -0.040 | -0.100 | -0.256 | -0.171 | -0.002 | 0.138 | 0.126 | 0.328 | -0.340 | -0.136 |
| 329 | 9877 | 0.237 | 0.110 | 0.099 | 0.052 | -0.070 | 0.088 | 0.311 | 0.133 | -0.094 | -0.326 | -0.258 | -0.066 | -0.190 | -0.029 |
| 445 | 13412 | -0.172 | -0.200 | -0.177 | -0.222 | -0.020 | -0.105 | -0.219 | -0.184 | -0.107 | 0.222 | 0.011 | 1.626 | -0.279 | -0.166 |
| 405 | 12151 | -0.342 | -0.457 | -0.509 | -0.076 | 0.304 | 0.656 | -0.031 | 0.355 | -0.117 | 0.647 | 0.220 | 0.076 | -0.268 | -0.461 |
| 19 | 710 | -0.510 | -0.318 | -0.391 | -0.206 | 0.275 | 0.155 | ··· | -0.230 | ··· | 0.439 | 0.423 | 0.332 | 0.022 | 0.005 |
| 59 | 1894 | -0.351 | -0.207 | -0.246 | 0.357 | -0.093 | 0.748 | 0.161 | 0.153 | ··· | 0.338 | 0.623 | -0.151 | -0.668 | -0.658 |
| 208 | 6442 | 0.839 | 0.877 | 0.809 | 0.193 | -0.448 | -0.410 | -0.207 | -0.013 | -0.205 | -0.283 | -0.210 | -0.062 | -0.555 | -0.326 |
| 333 | 9934 | 0.475 | 0.439 | 0.340 | -0.131 | 0.033 | -0.282 | -0.705 | -0.509 | -0.307 | 0.010 | 0.267 | 0.059 | 0.111 | 0.198 |
| 497 | 15248 | -0.165 | -0.205 | -0.330 | -0.151 | -0.295 | -0.216 | -0.208 | -0.215 | -0.216 | -0.034 | -0.170 | 1.892 | 0.246 | 0.072 |
| 295 | 8766 | -0.192 | -0.231 | -0.234 | 1.158 | -0.002 | -0.151 | -0.287 | -0.297 | -0.042 | -0.080 | 0.950 | -0.270 | -0.059 | -0.254 |
| 338 | 10028 | -0.098 | -0.166 | -0.133 | 0.780 | -0.097 | -0.125 | 0.196 | -0.093 | -0.177 | 0.089 | 0.315 | -0.208 | -0.014 | -0.270 |
| 664 | 18996 | -0.003 | 0.073 | -0.015 | 0.324 | -0.070 | 0.244 | -0.275 | -0.119 | ··· | -0.156 | 0.048 | 0.439 | -0.305 | -0.178 |
| 638 | 18424 | -0.006 | 0.064 | 0.113 | 0.161 | 0.207 | 0.227 | 0.040 | 0.151 | ··· | -0.252 | -0.066 | 0.177 | -0.518 | -0.301 |
| 258 | 7756 | 0.204 | -0.270 | -0.274 | 0.153 | -0.169 | 0.026 | 0.129 | 0.250 | -0.125 | 0.206 | -0.087 | 0.467 | -0.322 | -0.184 |
| 667 | 19040 | 0.790 | 0.745 | 0.525 | 0.255 | 0.150 | -0.029 | 0.086 | 0.096 | -0.338 | -0.553 | -0.567 | -0.563 | -0.381 | -0.220 |
| 598 | 17475 | -0.541 | -0.560 | -0.550 | 0.294 | 1.454 | 0.858 | 0.565 | 0.194 | -0.134 | 0.008 | -0.208 | -0.838 | -0.243 | -0.303 |

TABLE 5

VARIABILITY PARAMETERS $A$ AND $T$

| Sample | $N_{obj}$ | $N_{points}$ | $A$ | | | | $T$ | | | |
|---|---|---|---|---|---|---|---|---|---|---|
| | | | | $\epsilon$(70%) | $\epsilon$(95%) | $\epsilon$(99%) | | $\epsilon$(70%) | $\epsilon$(95%) | $\epsilon$(99%) |
| all | 35 | all 14 | 0.159 | 0.009 | 0.017 | 0.022 | 0.976 | 0.128 | 0.242 | 0.319 |
| low-z | 18 | first 12 | 0.114 | 0.012 | 0.023 | 0.030 | 1.048 | 0.251 | 0.475 | 0.622 |
| high-z | 17 | first 12 | 0.194 | 0.017 | 0.032 | 0.042 | 1.034 | 0.186 | 0.352 | 0.461 |
| low-L | 17 | first 12 | 0.215 | 0.026 | 0.050 | 0.066 | 1.463 | 0.344 | 0.648 | 0.847 |
| high-L | 18 | first 12 | 0.109 | 0.008 | 0.015 | 0.020 | 0.630 | 0.111 | 0.210 | 0.276 |

## Figure Captions

FIG. 1. – Weighted root-mean-square deviation of the magnitude differences with respect to the mean for each object, over the 15 year baseline, as a function of J magnitude ($m_4$(1053)) for the 694 stellar objects in the survey. The variability criterion is indicated by the dotted line and is derived from the photometric errors (Table 2). Squares: spectroscopically confirmed quasars. Circled: variable objects with proper motion greater than four times the one-sigma errors. Triangles: compact objects with narrow emission lines (to show that all but one are not detected to vary).

FIG. 2a. – Proper motion, in units of arcsec/century, toward the Galactic center ($\mu_0$) and in the orthogonal direction ($\mu_{90}$) for the spectroscopically confirmed objects. Small, filled symbols: the 35 quasars. Large, open symbols: the 6 stars.

FIG. 2b. – Same as for Fig 2a, but for the 20 objects without spectroscopic identification that have proper motion measurements. The filled symbols indicate objects in the KKC list, i.e., those that have colors unlike normal stars.

FIG. 3. – $U - J$ versus $J - F$ diagrams for all of the objects. Comparison can be made directly to similar diagrams in KKC and T89. Left: $19.0 < J_K < 21.5$ (note that a few variables are brighter than this interval). Middle: $21.5 < J_K < 22.5$. Right: $J_K > 22.5$. Filled circles indicate spectroscopically confirmed variables and open circles indicate spectroscopically unconfirmed variables. Dots are the non-variables (as in Fig. 1). Crosses are objects with stellar images but galaxy-like spectra (i.e., narrow lines). The small circles are the objects with $pmi < 4.0$ (excluding objects with galaxy-like spectra). For the others, the medium-sized circles indicate objects with proper motion $\mu < 1.0$ arcsec/century, and the large circles indicate objects with $\mu > 1.0$ arcsec/century. The three variables that fall just fainter than the KKC selection limit at $J_K = 22.5$ are indicated by squares in the right-hand panel. The one non-variable quasar is indicated as stellated point in the center panel.

FIG. 4a. – The apparent (observed-frame) square of the structure function for the 35 known quasars. Each point represents a plate pair, and the error bars indicate the r.m.s. dispersion over the ensemble of quasars. The contribution from photometric noise can be inferred from the value of the structure function at zero time lag (derived from plate pairs at the same epoch).

FIG. 4b. – Same as for Fig. 4a, for the 15 variables with $pmi \geq 4.0$. Here the points are binned in 1-year intervals, and the error bars indicate the error in the mean of the data in the bin.

FIG. 4c. – Same as for Fig. 4b, for the variables without detected proper motion and without spectroscopic confirmation. Of the 10 such variables, the faintest 4 have been omitted in order to make the distribution of apparent magnitudes for this sample more nearly comparable to that of Fig. 4a and Fig. 4b.

FIG. 5a. – Rest-frame structure function for the known quasars. The last point contains at least 65% of the total sample (23 out of 35 objects in this case), a criterion used throughout these plots. The fitting curve is described in the text. The horizontal line gives $2 < \delta^2 >$ (see text).

FIG. 5b. – Rest-frame structure function for the 18 quasars of lowest redshift, $z < 1.34$.

FIG. 5c. – Rest-frame structure function for the 17 quasars of highest redshift, $z > 1.34$.

FIG. 5d. – Rest-frame structure function for the 17 quasars of lowest luminosity, $M_J > -23.92$.

FIG. 5e. – Rest-frame structure function for the 18 quasars of highest luminosity, $M_J < -23.92$.

FIG. 6a. – Correlation function for the same sample as in Figure 5a.

FIG. 6b. – Correlation function for the same sample as in Figure 5b.

FIG. 6c. – Correlation function for the same sample as in Figure 5c.

FIG. 6d. – Correlation function for the same sample as in Figure 5d.

FIG. 6e. – Correlation function for the same sample as in Figure 5e.

FIG. 7. – Distribution of the 35 quasars in log $z$ versus $M_J$. ($H_0 = 50$ km sec$^{-1}$ Mpc$^{-1}$, $q_0 = 0.5$, and $f_\nu \sim \nu^\alpha$ with $\alpha = -1$.) The diagonal line gives the apparent magnitude limit $J = 22.5$, and the vertical and horizontal lines are at the median luminosity ($M_J = -23.92$) and redshift (1.34), respectively.

FIG. 8a. – Correlation function in the rest-frame for the half of the sample with $\xi > -24.4$, where $\xi = M_J - 5\log f(z)$, where $f(z) = 2[(1+z) - \sqrt{(1+z)}]$.

FIG. 8b. – As for Fig. 8a, but for the half of the sample with $\xi < -24.4$.

FIG. 8c. – Structure function in the rest-frame for the half of the sample with $\xi > -24.4$.

FIG. 8d. – Structure function in the rest-frame for the half of the sample with $\xi < -24.4$.

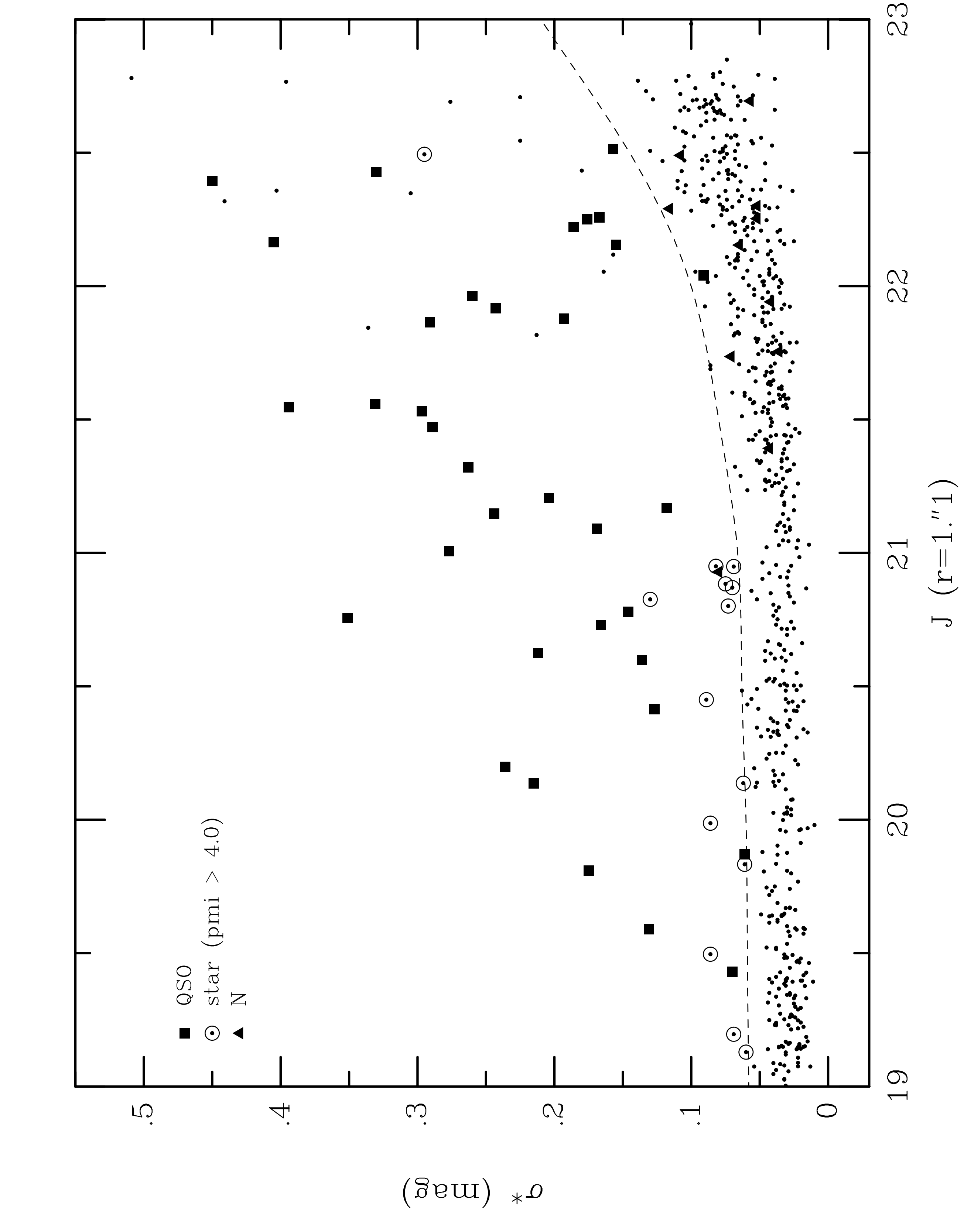

Fig. 1

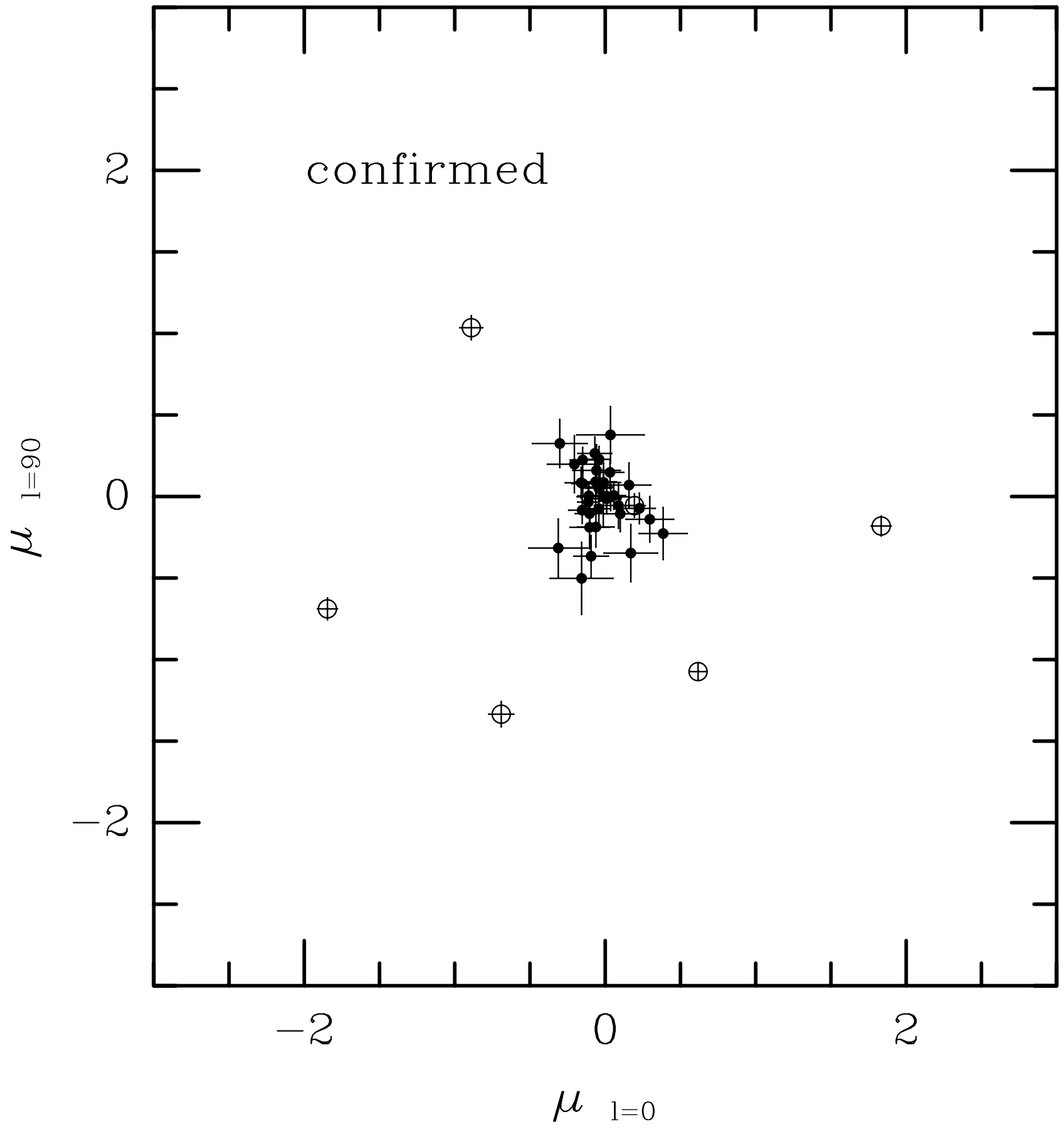

Fig. 2a

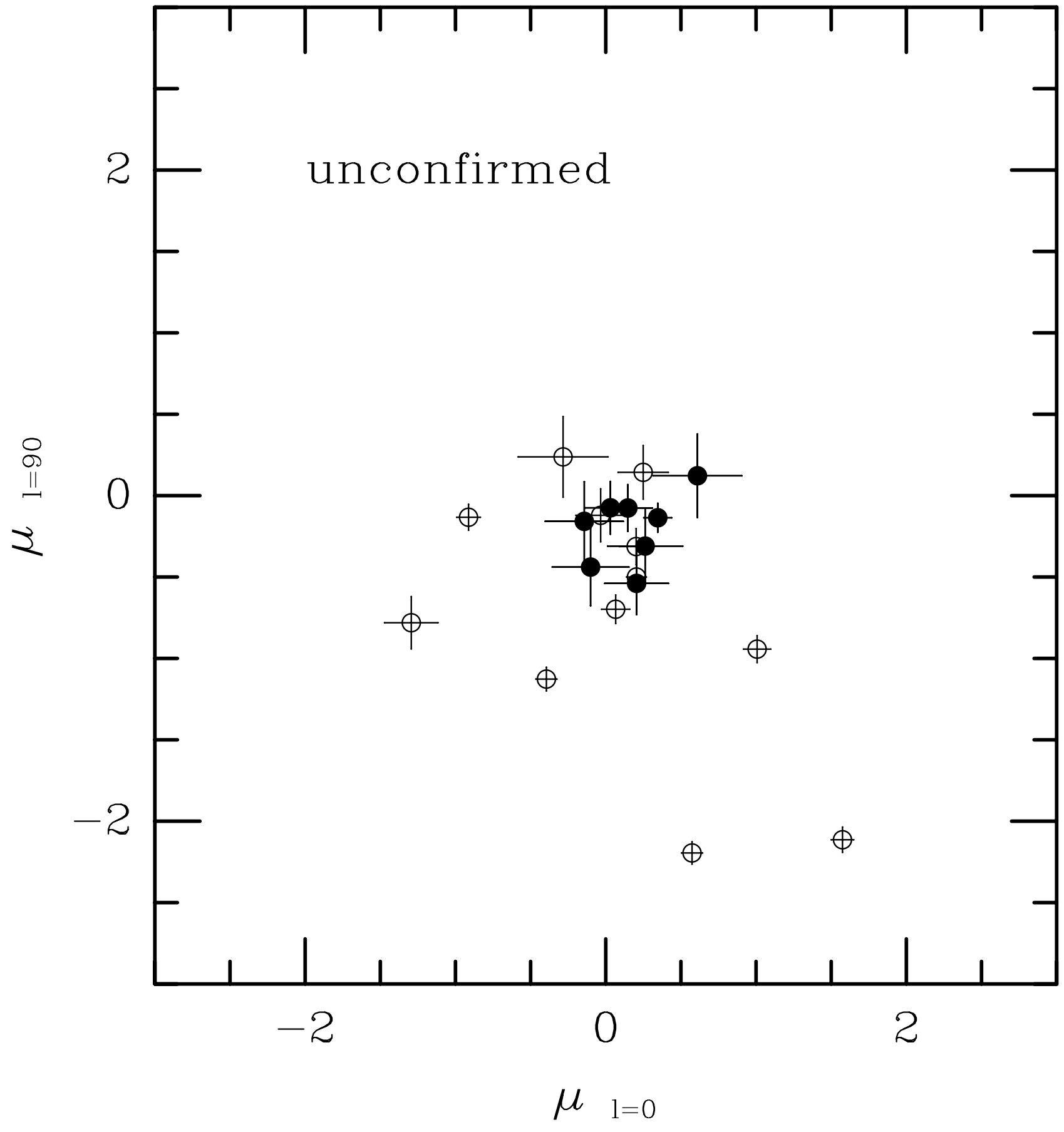

Fig. 2b

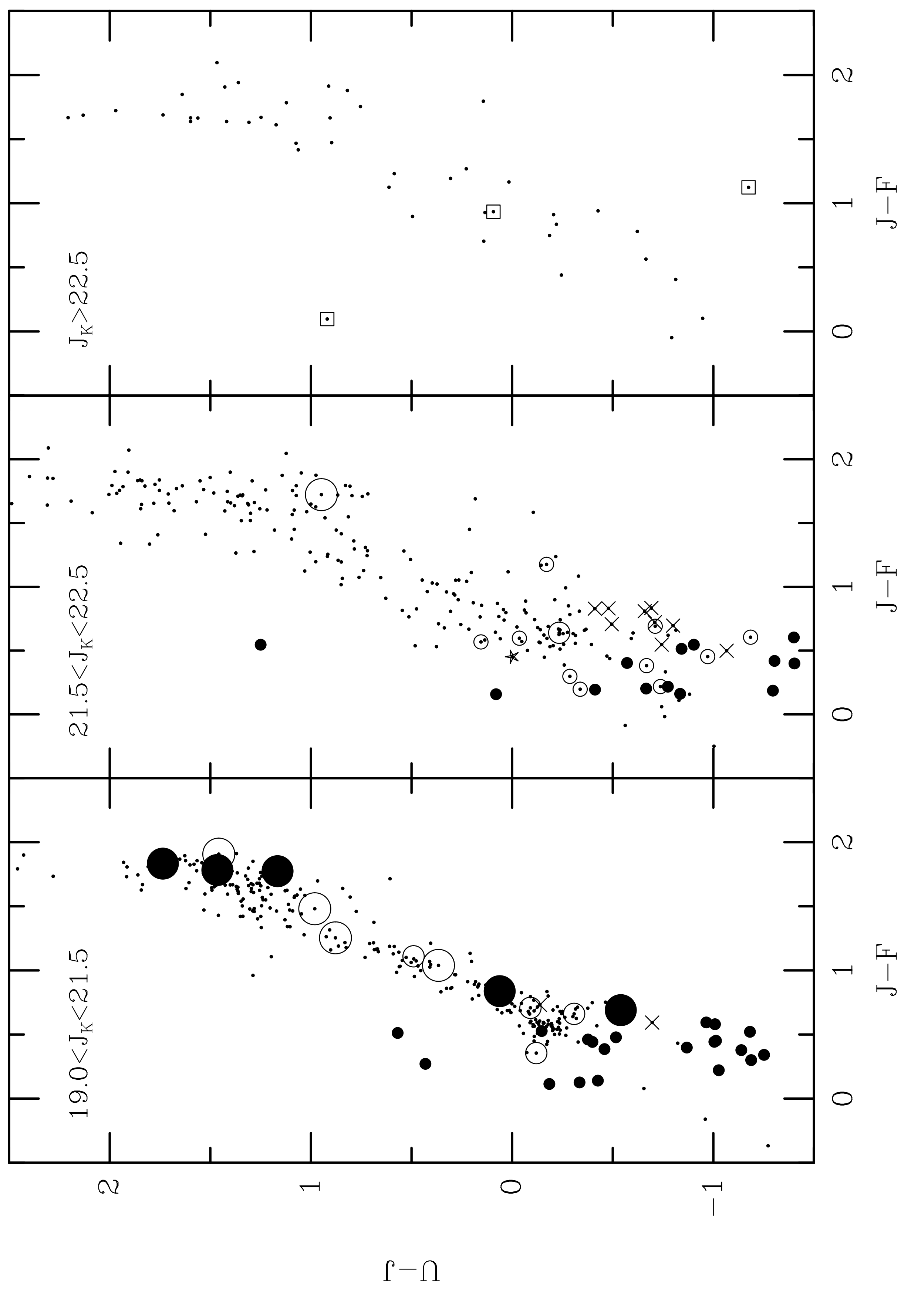

Fig. 3

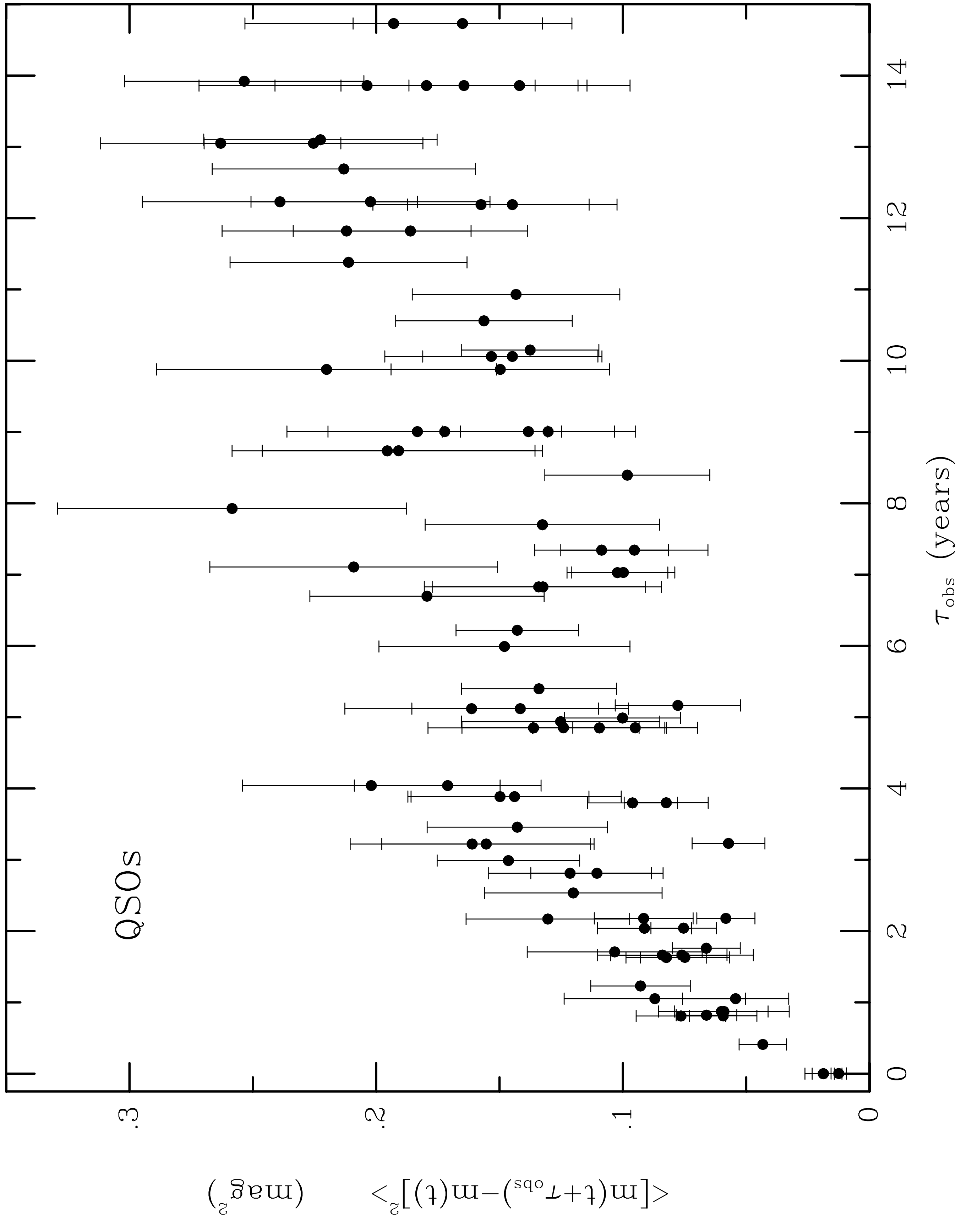

Fig. 4a

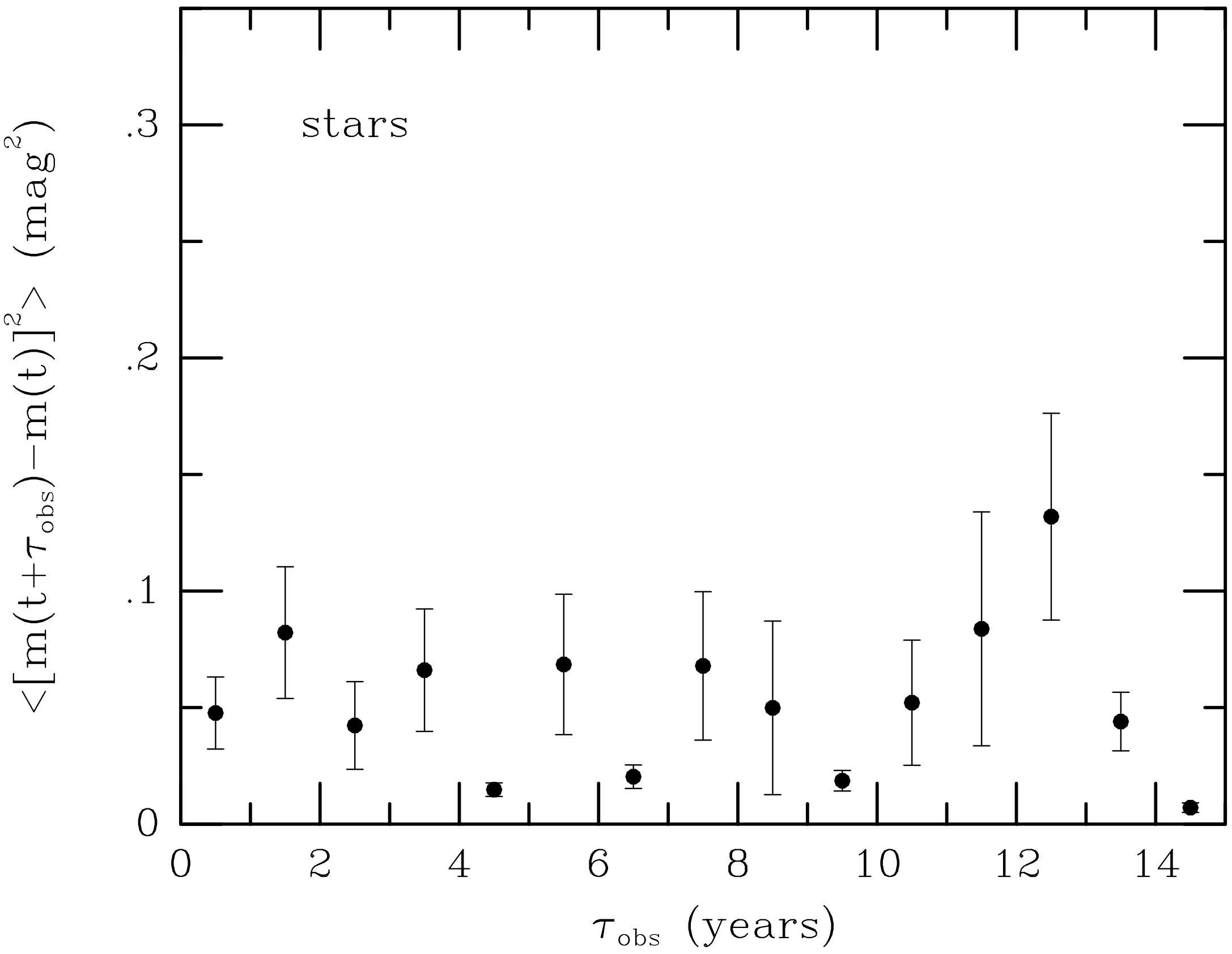

Fig. 4b

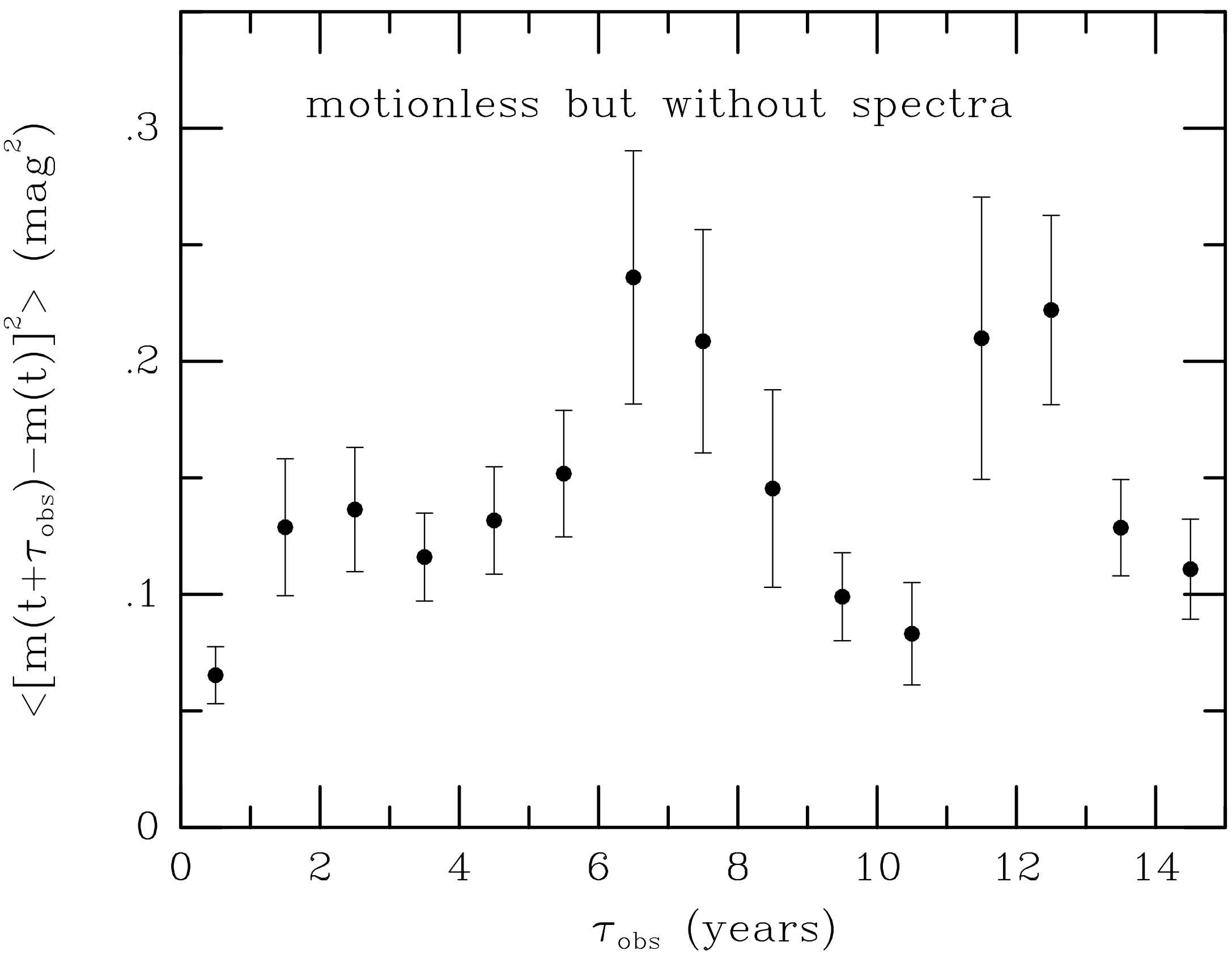

Fig. 4c

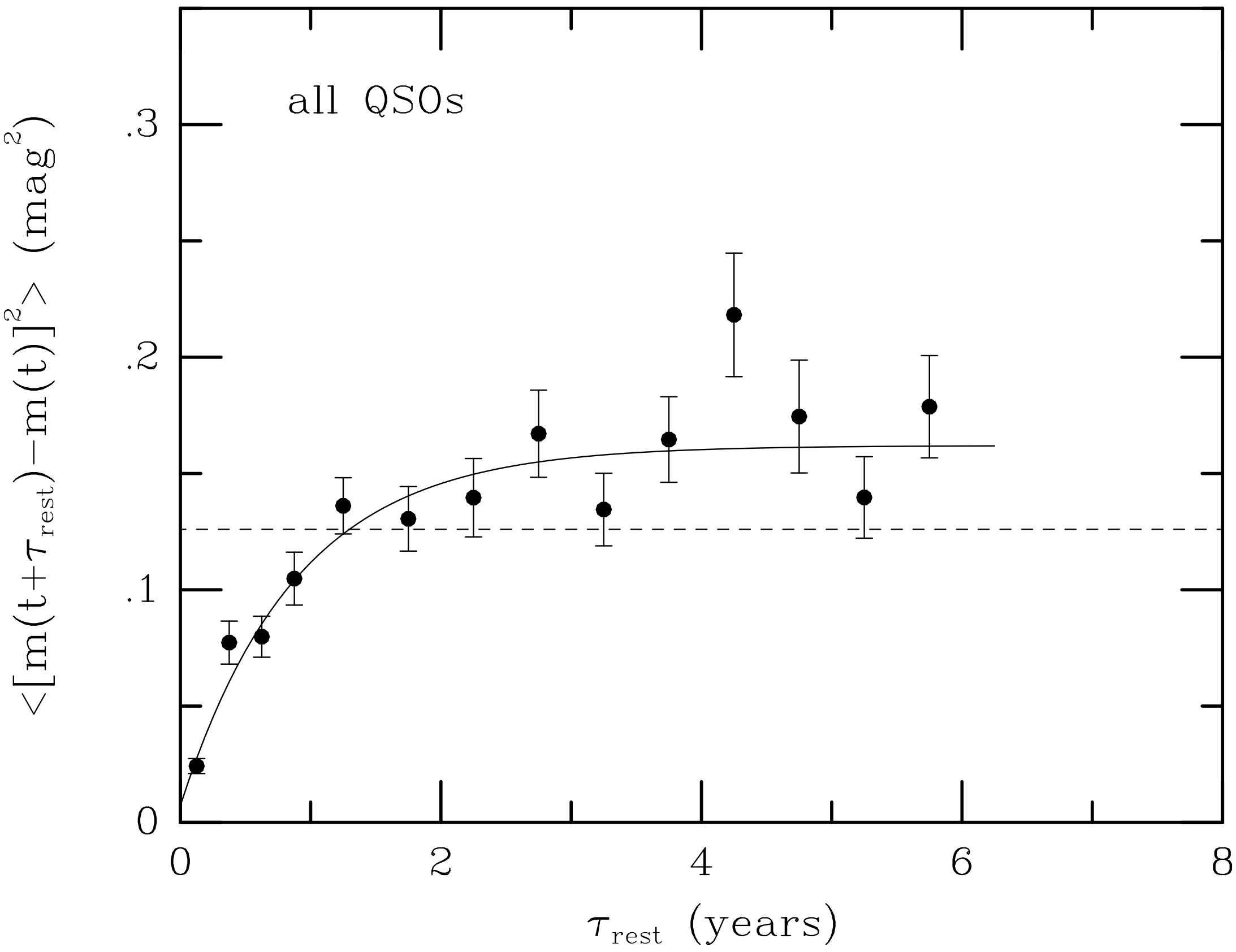

Fig. 5a

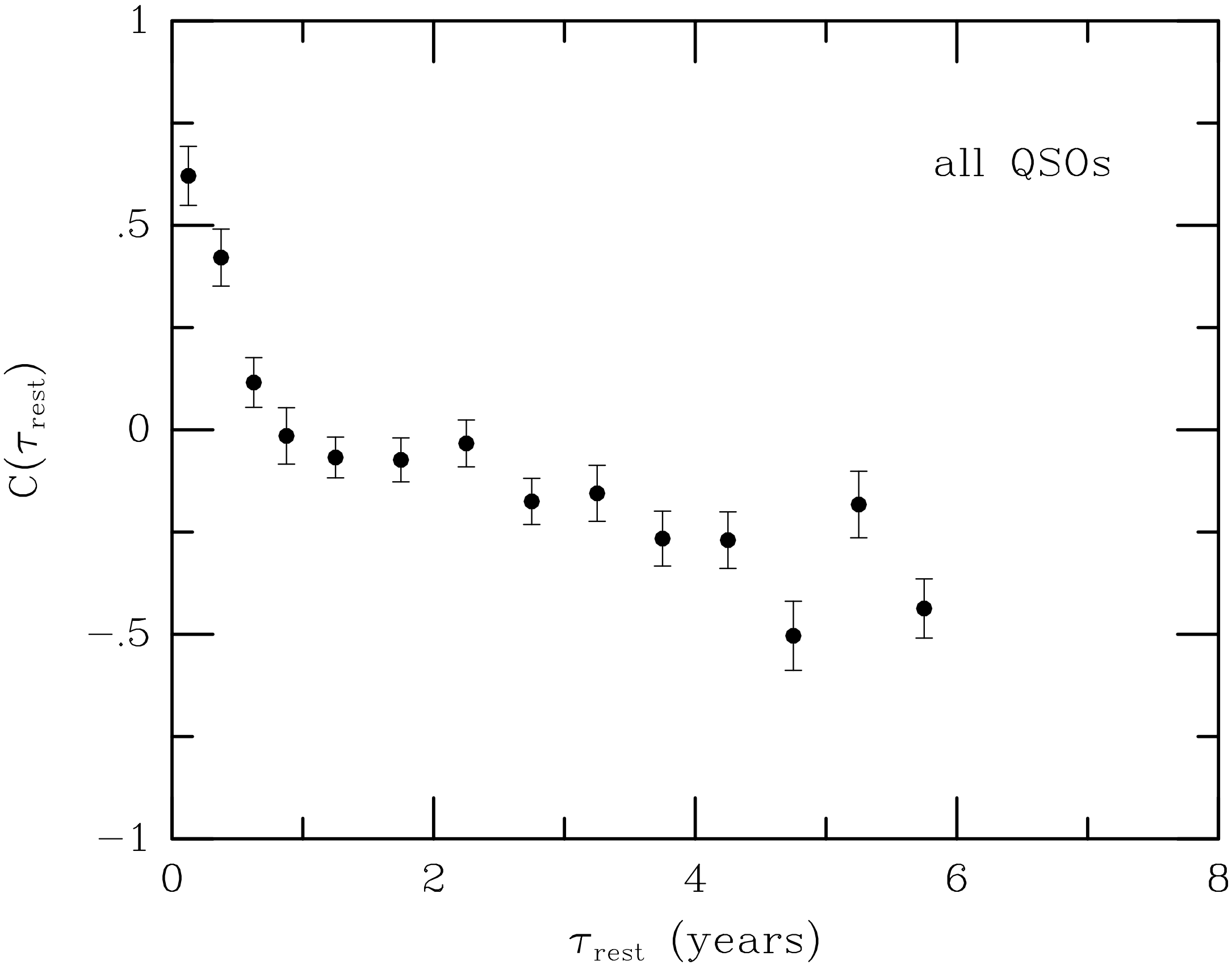

Fig. 6a

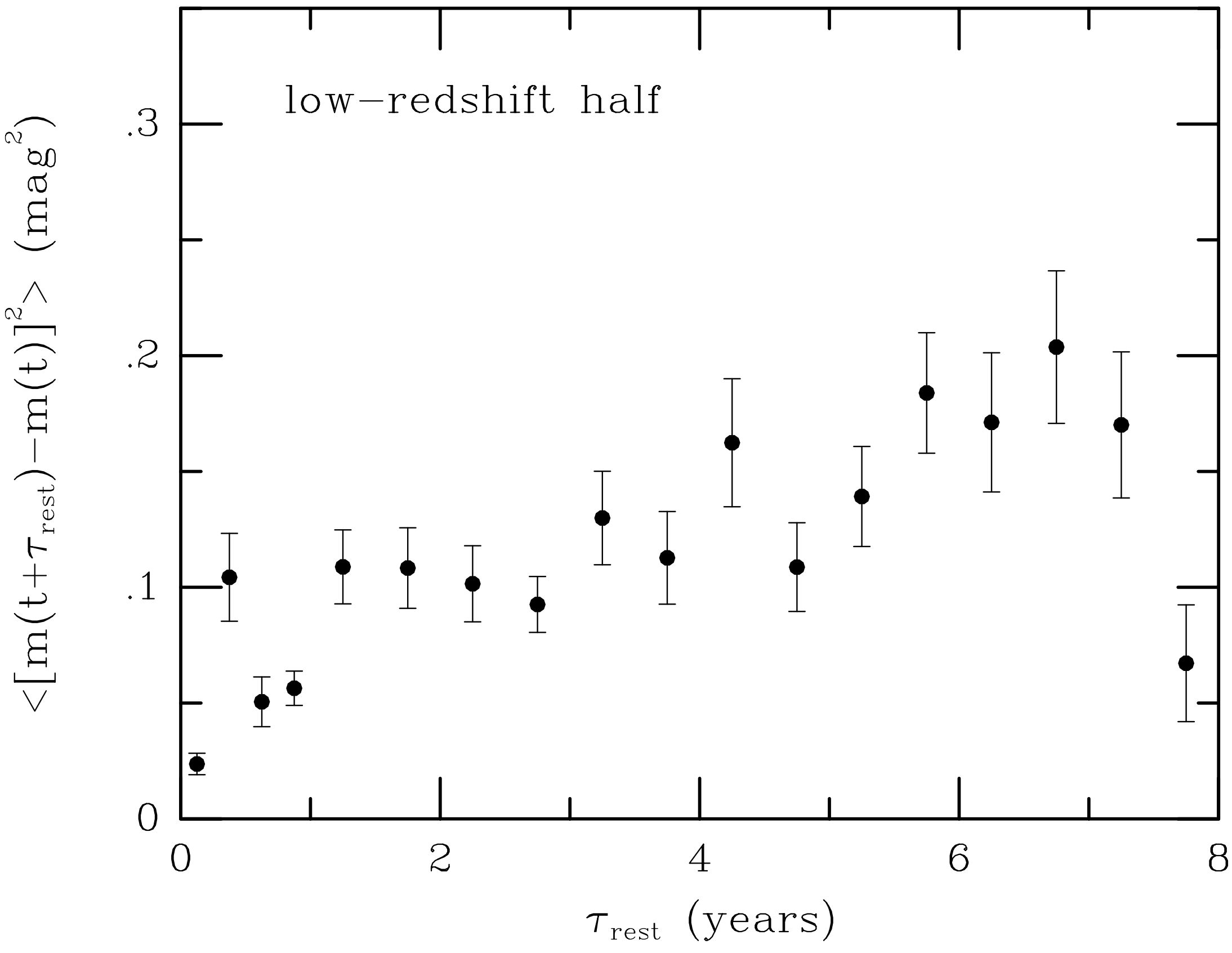

Fig. 5b

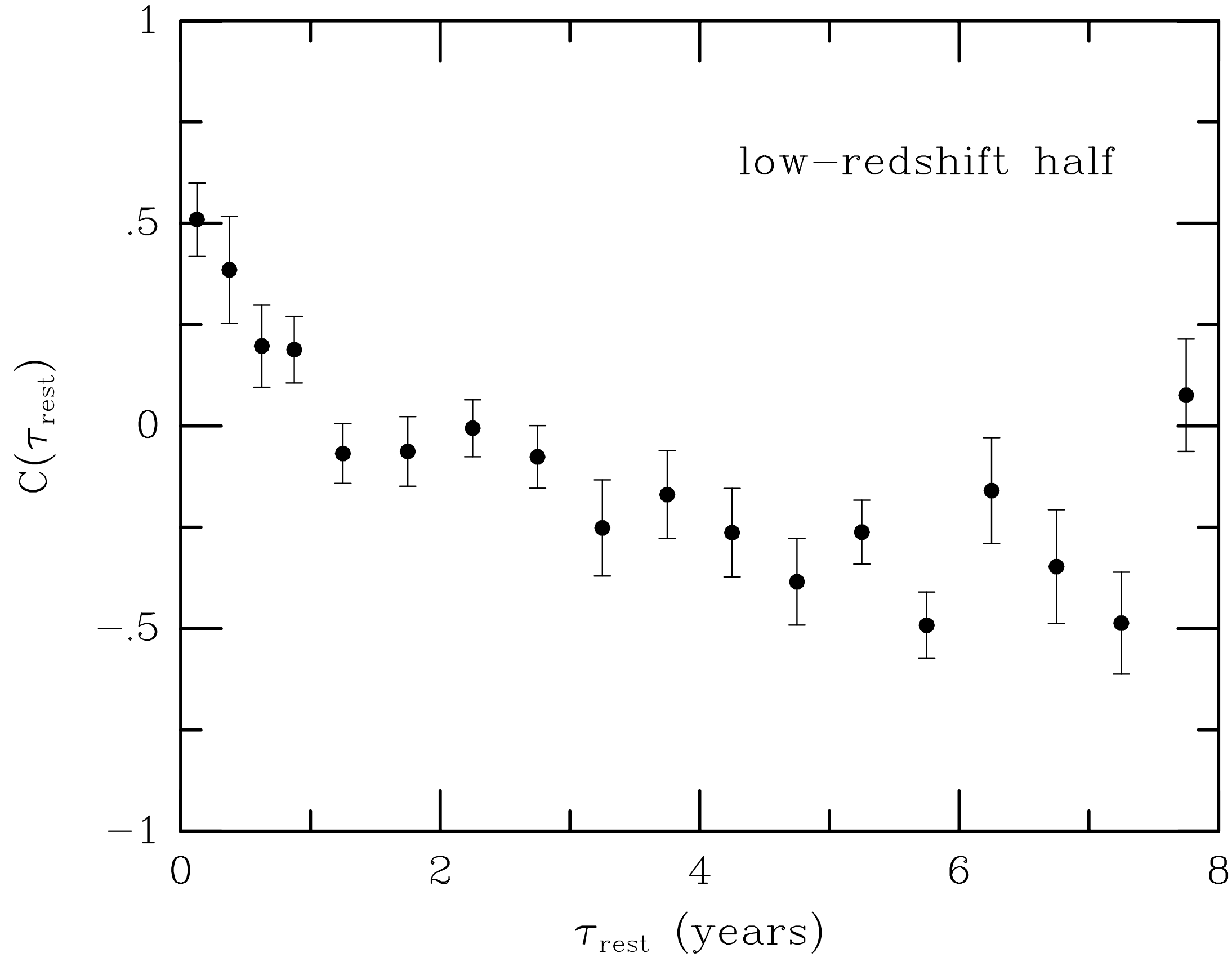

Fig. 6b

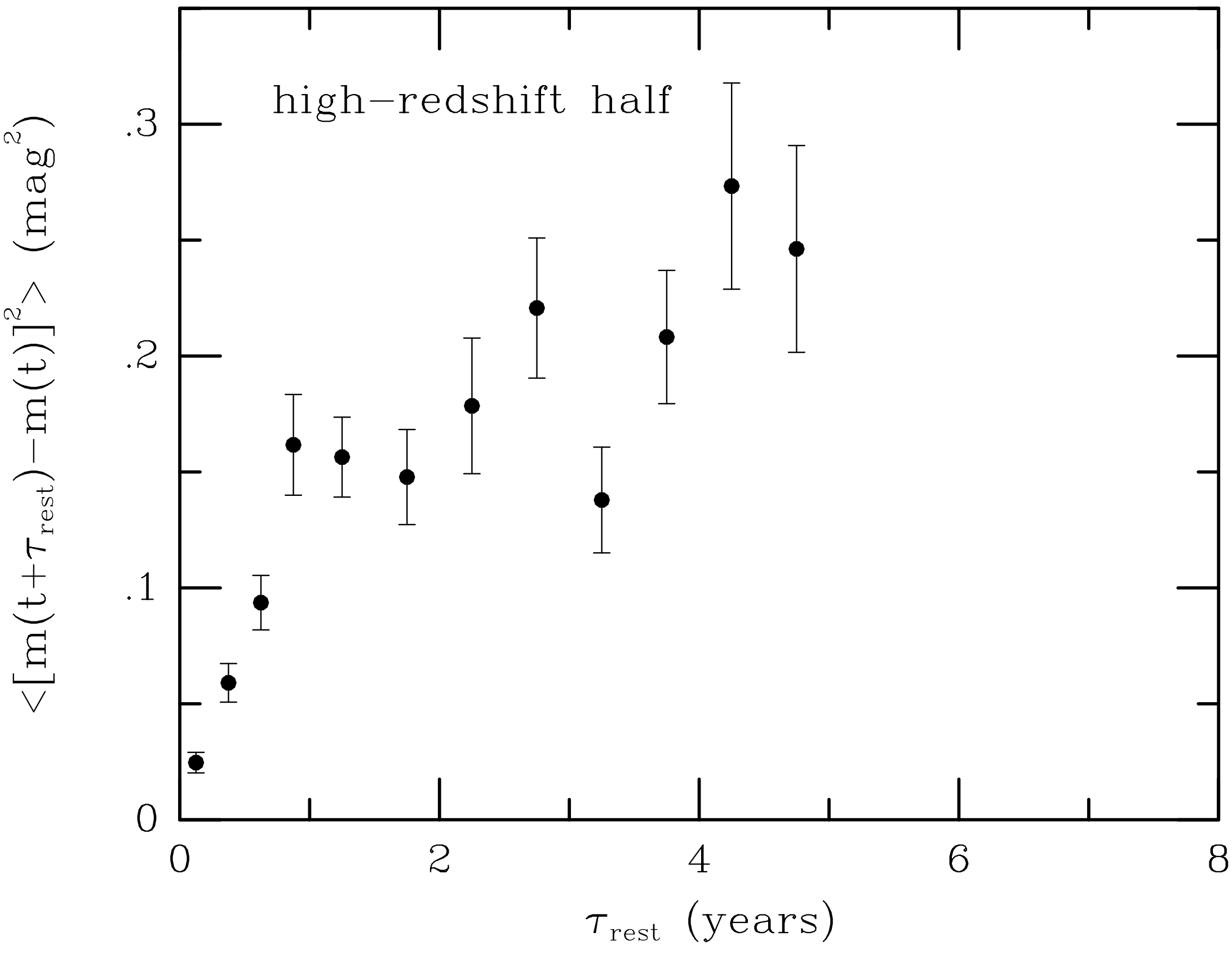

Fig. 5c

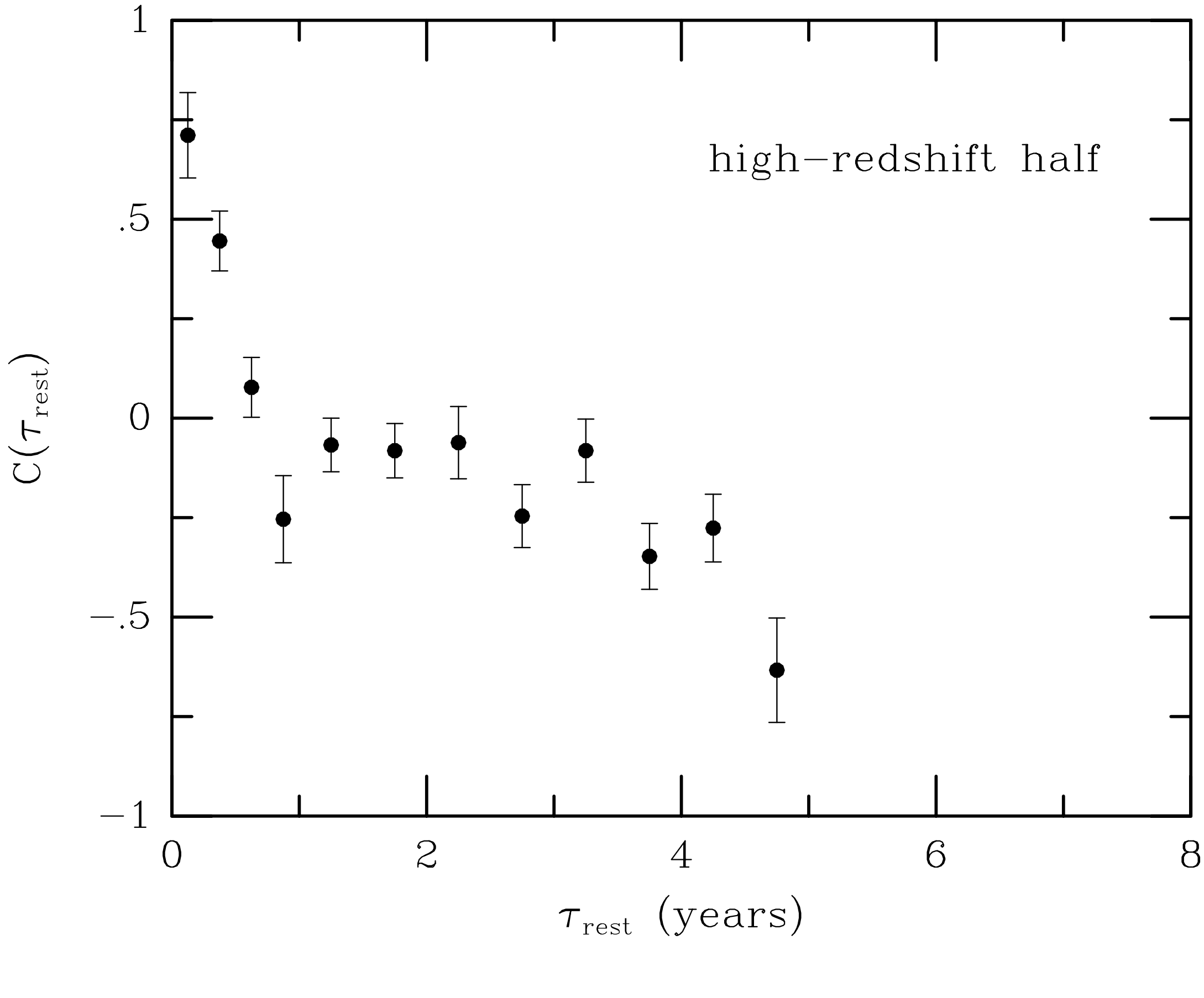

Fig. 6c

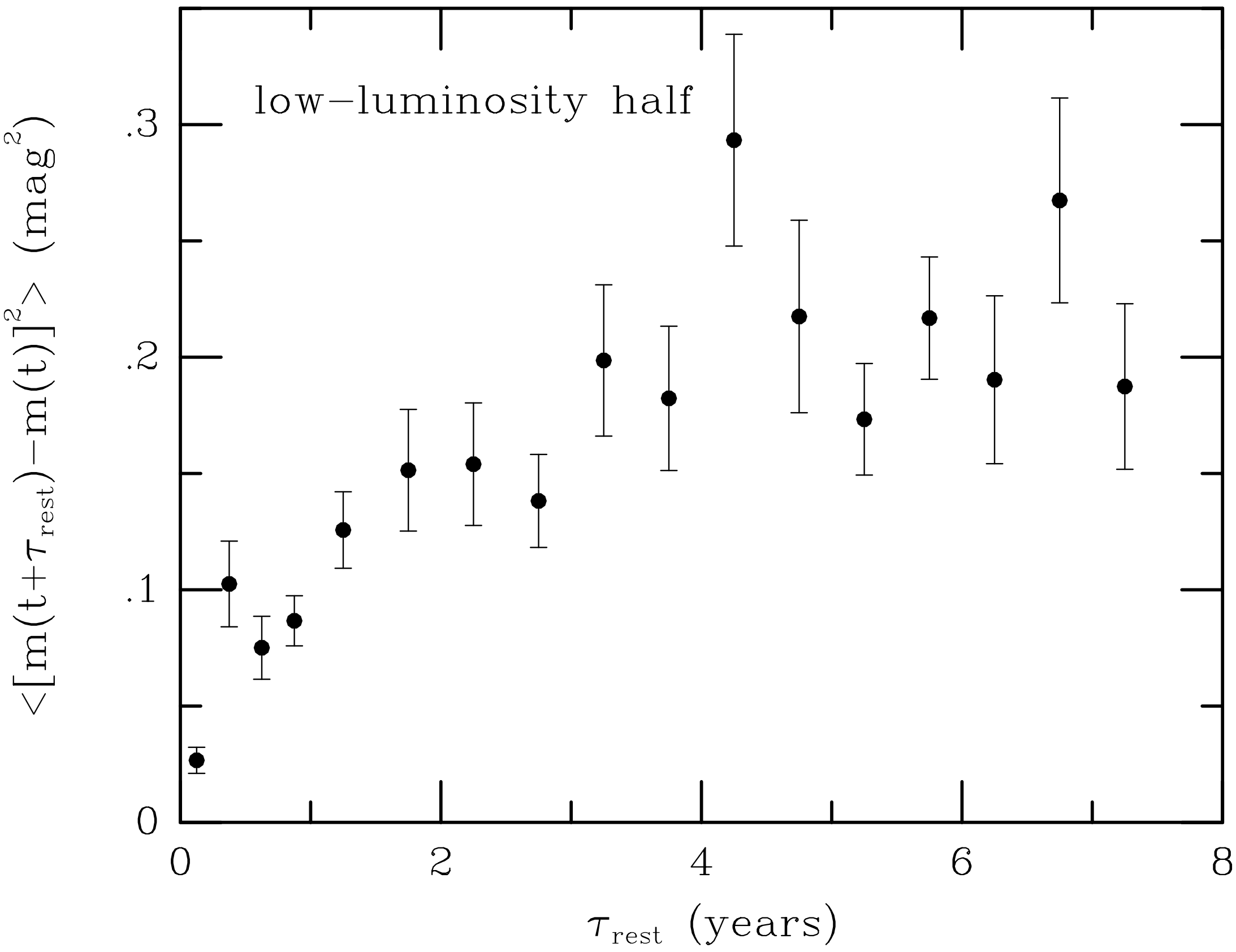

Fig. 5d

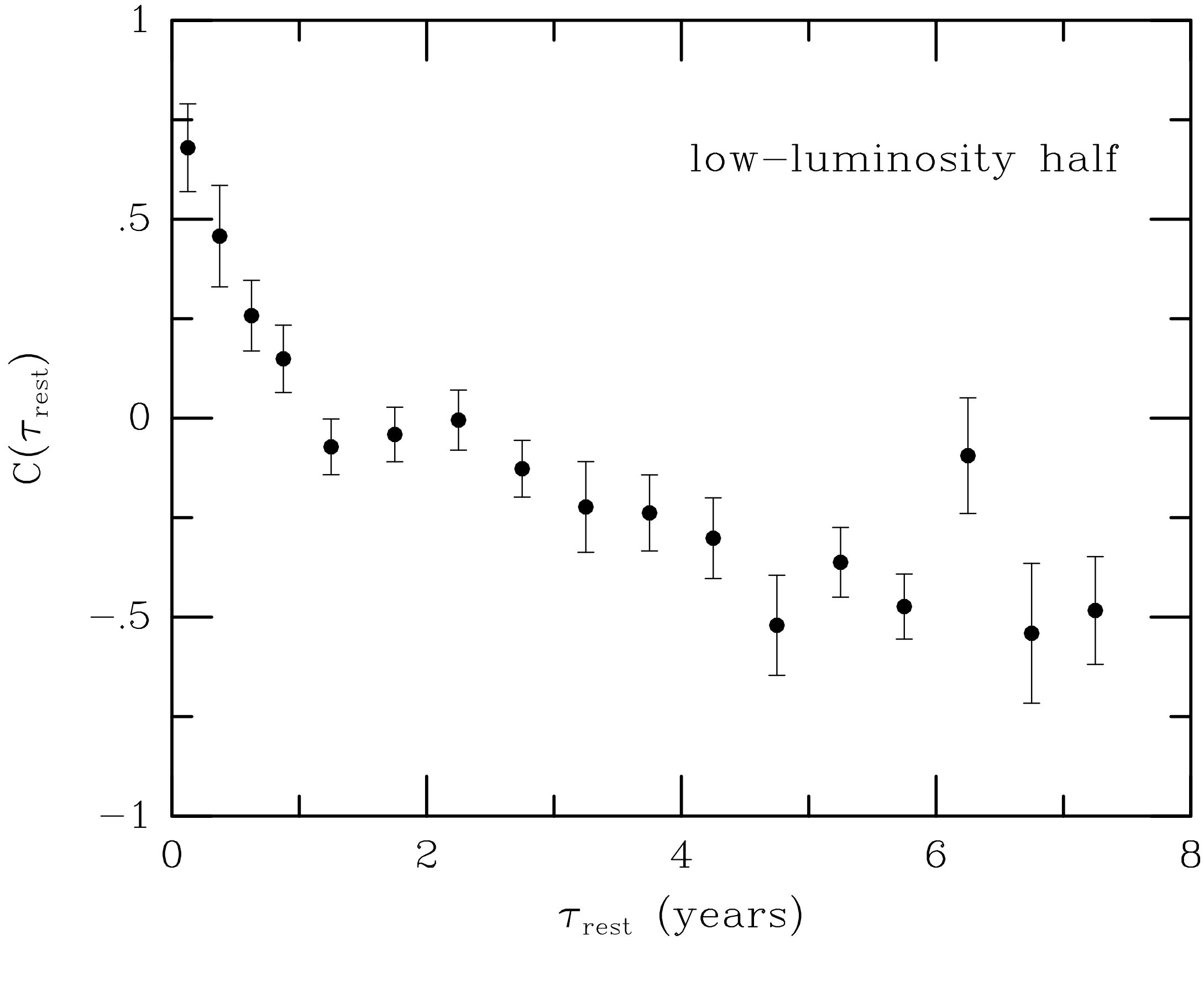

Fig. 6d

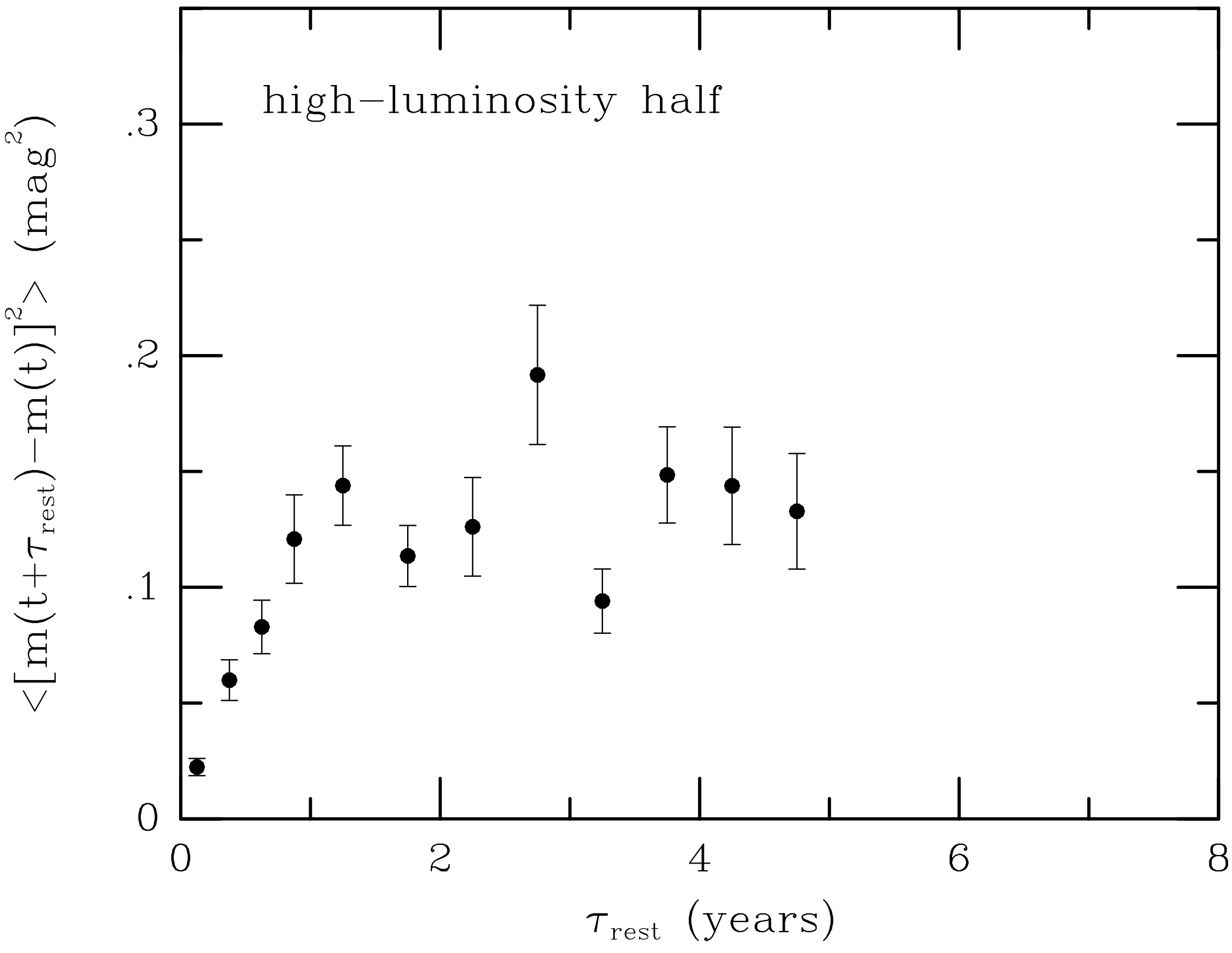

Fig. 5e

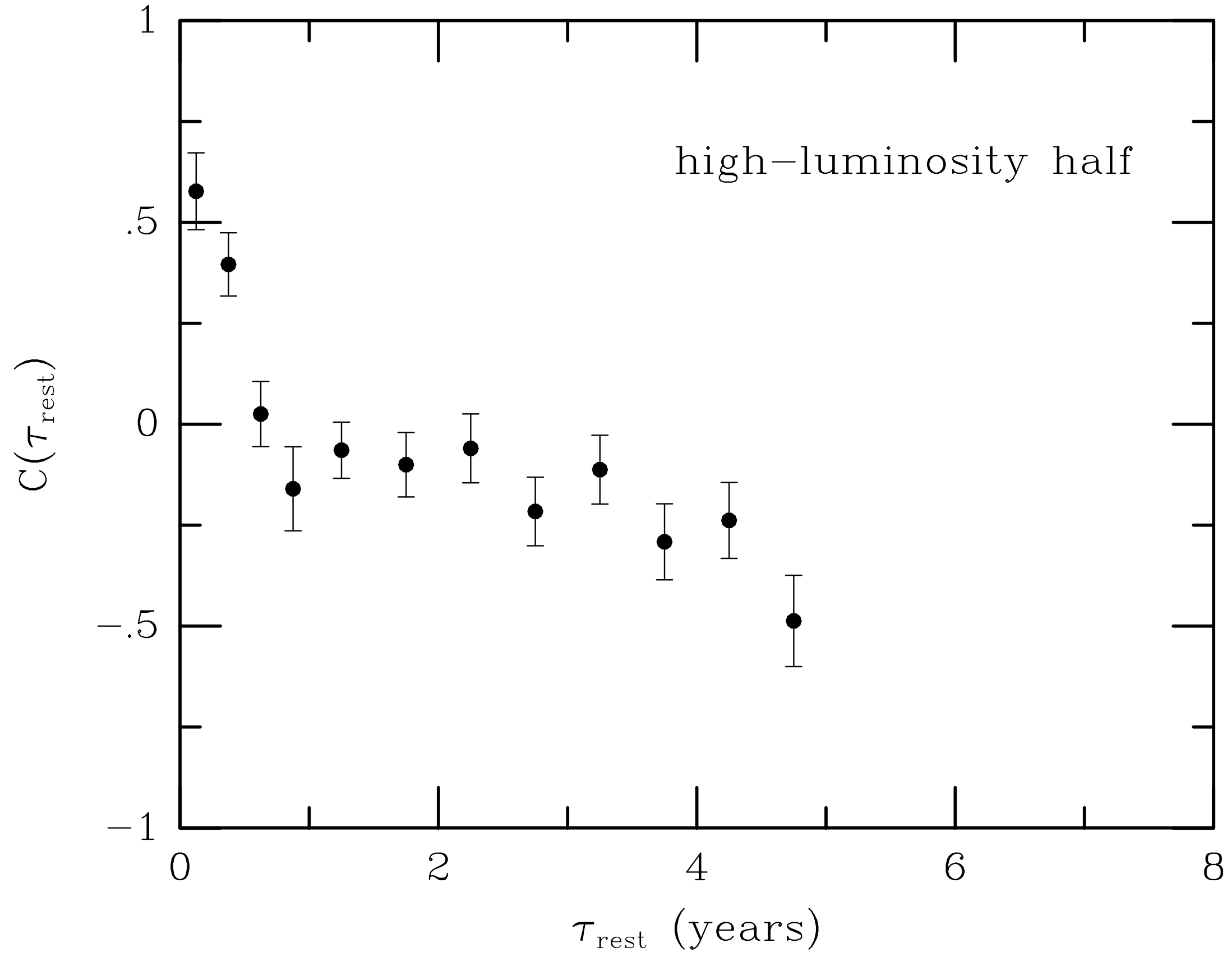

Fig. 6e

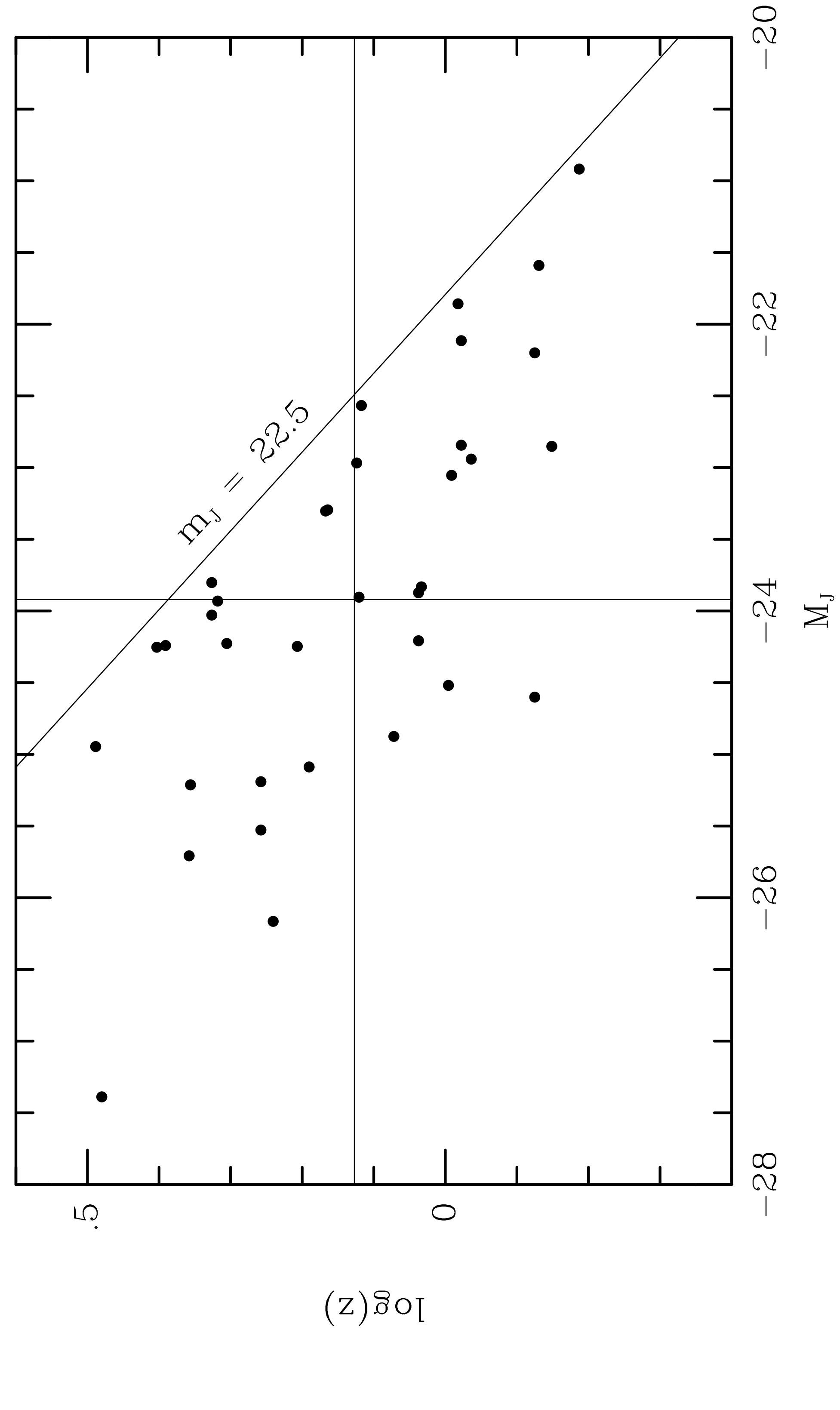

Fig. 7

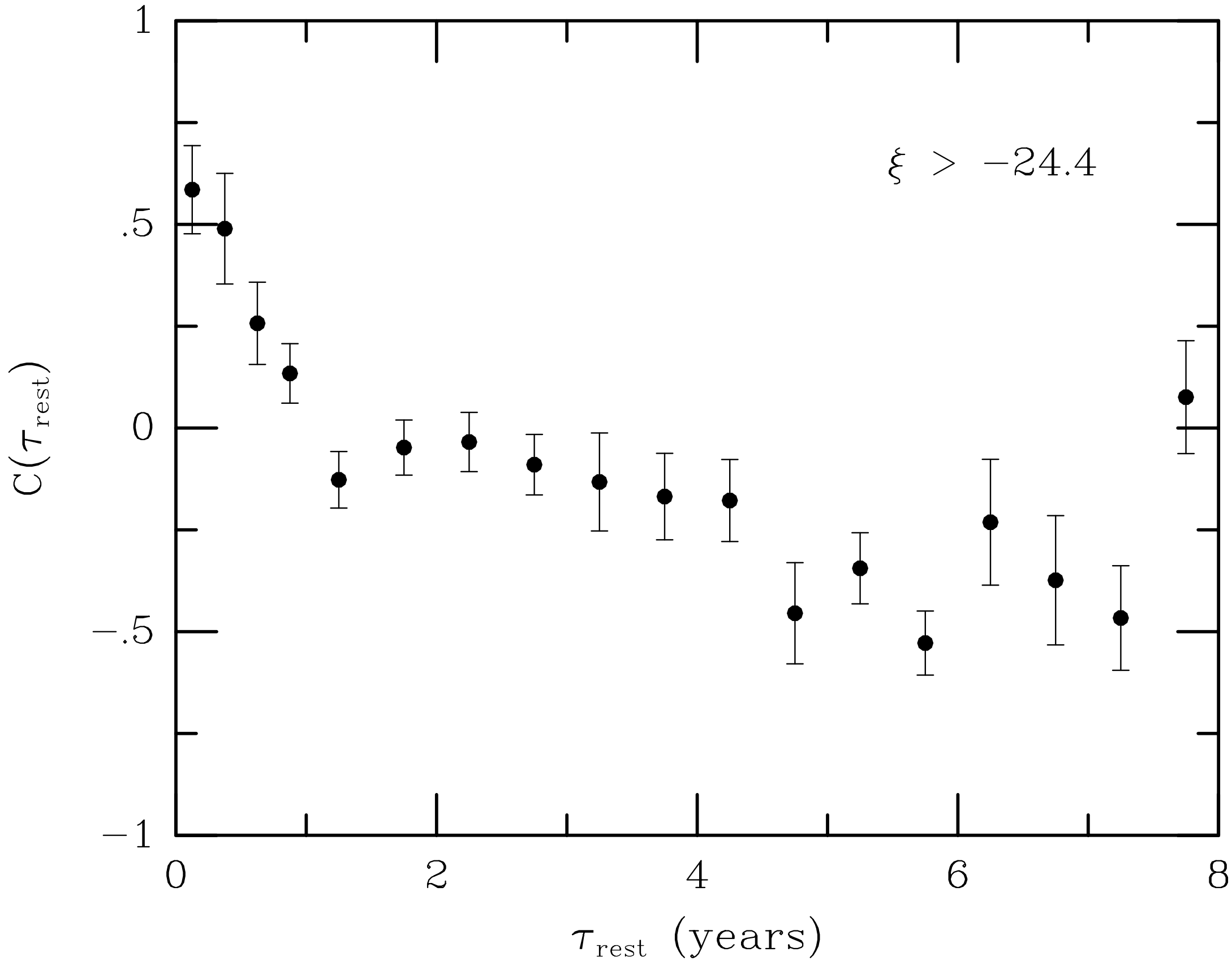

Fig. 8a

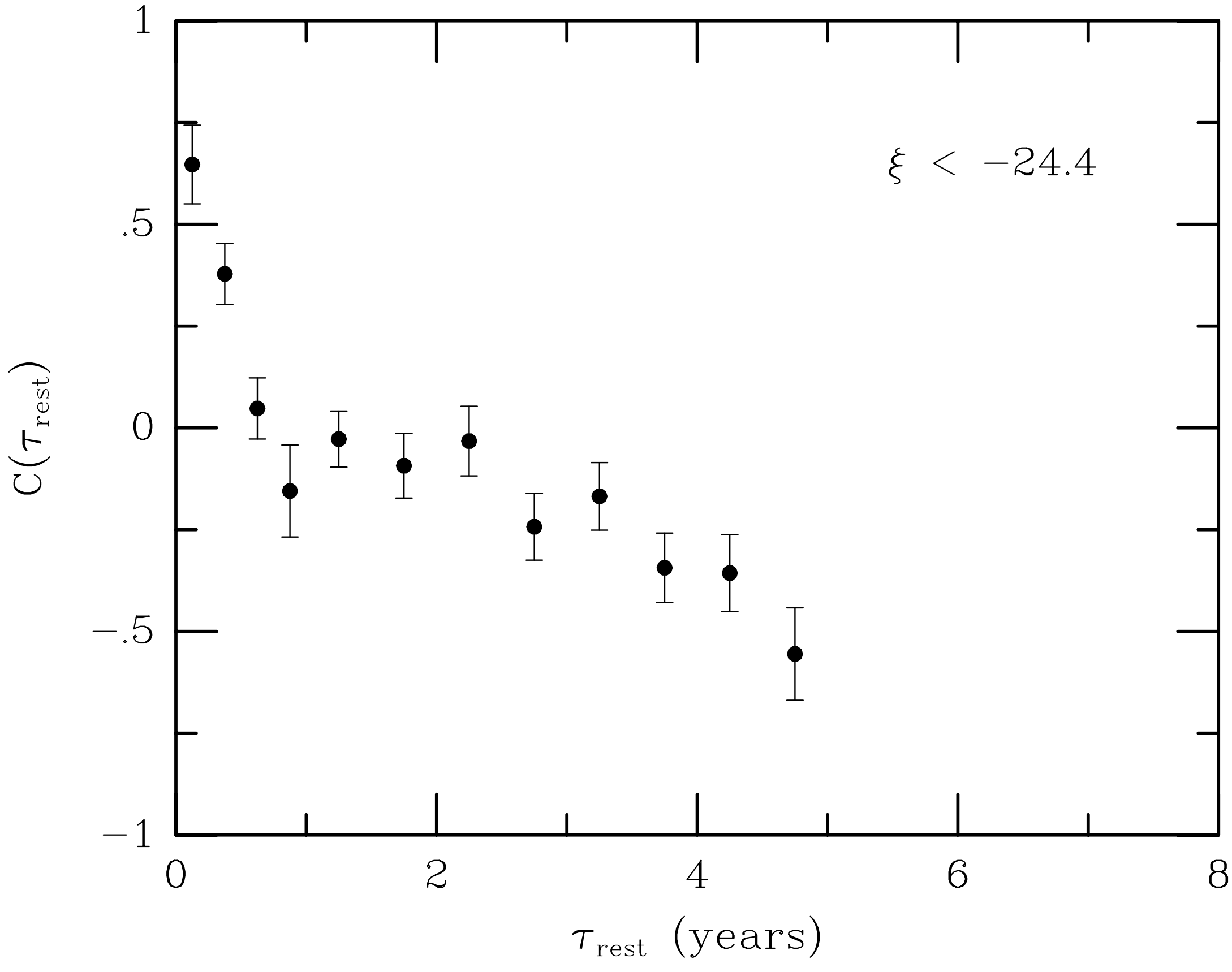

Fig. 8b

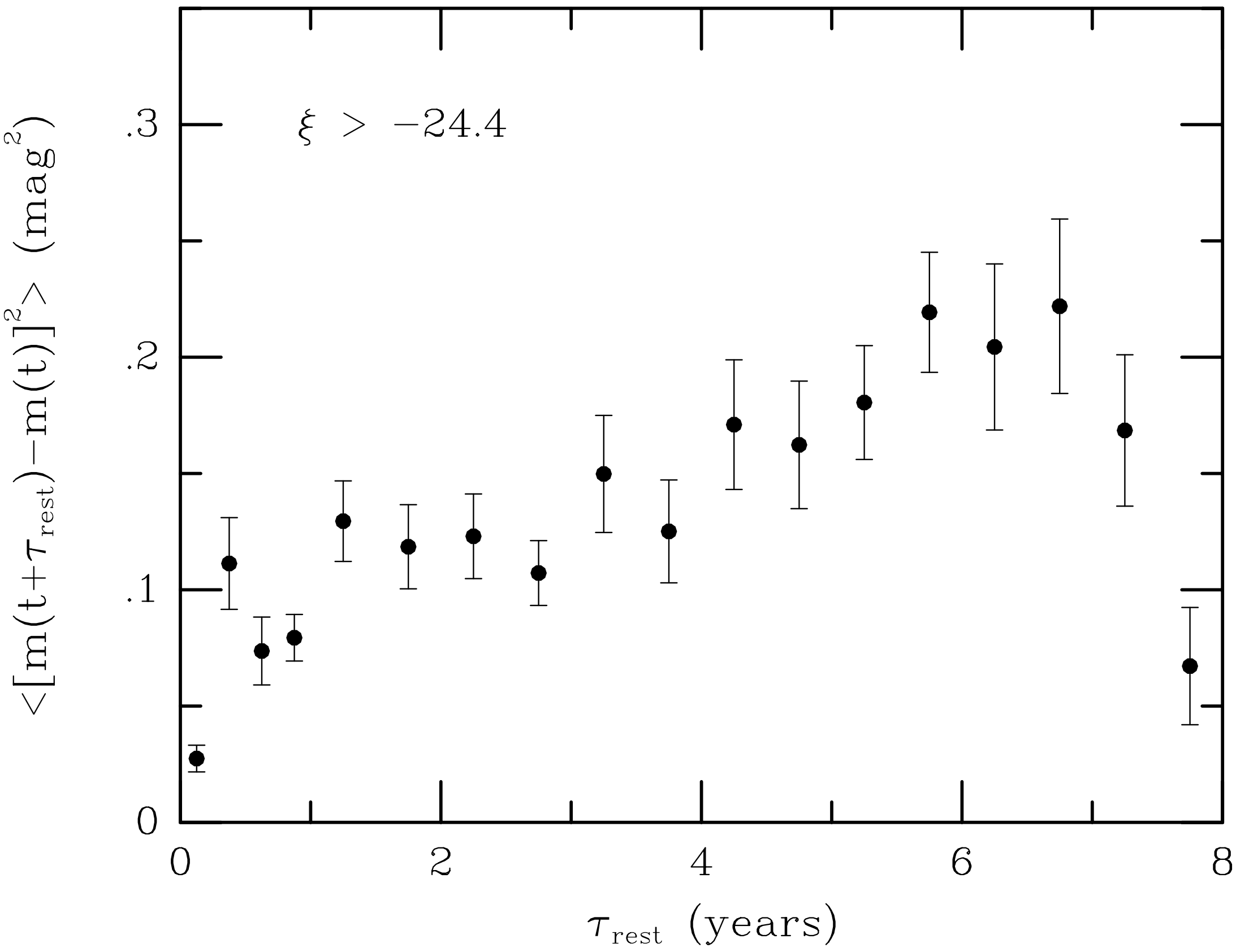

Fig. 8c

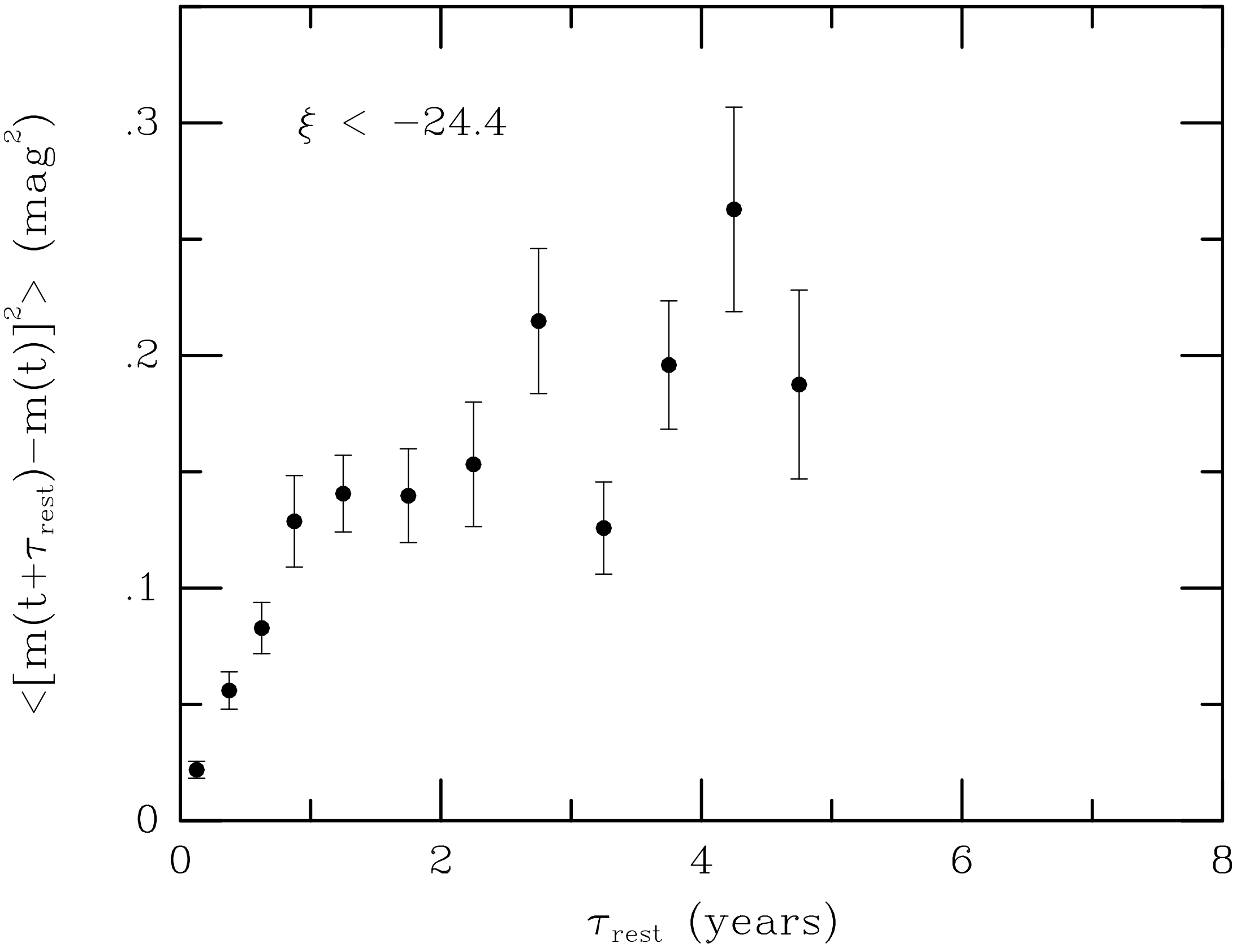

Fig. 8d